\documentclass[aps]{revtex4}
\usepackage{graphicx}
\usepackage{natbib}
\usepackage{latexsym}

\newcommand{\like}{\mathcal{L}}

\begin{document}

\title{Cosmic microwave background constraints on dark 
energy dynamics: analysis beyond the power spectrum}
\author{Fabio Giovi$^{1,2}$, Carlo Baccigalupi$^{1,2}$, Francesca
  Perrotta$^{1,2}$}
\address{$^{1}$SISSA/ISAS, Via Beirut 4, 34014 Trieste, Italy}
\address{$^{2}$INFN, Via Valerio 2, 34127 Trieste, Italy}

\begin{abstract}
We consider the distribution of the non-Gaussian signal induced by
weak lensing on the primary total intensity cosmic microwave
background (CMB) anisotropies. Our study focuses on the three point
statistics exploiting an harmonic analysis based on the CMB
bispectrum. By considering the three multipoles as independent
variables, we reveal a complex structure of peaks and valleys
determined by the re-projection of the primordial acoustic
oscillations through the lensing mechanism. We study the dependence of
this system on the expansion rate at the epoch in which the weak
lensing power injection is relevant, probing the dark energy equation
of state at redshift corresponding to the equivalence with matter or
higher ($w_\infty$). The variation of the latter quantity induces a
geometrical feature affecting distances and growth rate of linear
perturbations, acting coherently on the whole set of bispectrum
coefficients, regardless of the configuration of the three
multipoles. We evaluate the impact of the bispectrum observable on the
CMB capability of constraining the dark energy dynamics. We perform a
maximum likelihood analysis by varying the dark energy abundance, the
present equation of state $w_0$ and $w_\infty$. We show that the
projection degeneracy affecting a pure power spectrum analysis in
total intensity is broken if the bispectrum is taken into account. For
a Planck-like experiment, assuming nominal performance, no foregrounds
or systematics, and fixing all the parameters except $w_0$, $w_\infty$
and the dark energy abundance, a percent and ten percent precision
measure of $w_0$ and $w_\infty$ is achievable from CMB data only. The
reason is the enhanced sensitivity of the weak lensing signal to the
behavior of the dark energy at high redshifts, which compensates the
reduced signal to noise ratio with respect to the primary
anisotropies. These results indicate that the detection of the weak
lensing signal by the forthcoming CMB probes may be relevant to gain
insight into the dark energy dynamics at the onset of cosmic
acceleration.
\end{abstract}

\maketitle

\section{INTRODUCTION}
\label{intro}
Explaining the cosmic acceleration represents one of the greatest
challenges of modern cosmology. A reliable candidate for the dark
energy, a vacuum energy density similar to the Cosmological Constant
and responsible for the acceleration, should provide an answer to the
fine-tuning and the coincidence problem. The fine-tuning is required
to adjust the vacuum energy to the observed level, about 123 orders of
magnitude lower than the Planck energy scale. The coincidence question
is simply why the observed vacuum energy is comparable to the critical
energy density today (see \cite{lambda_review} and references
therein).

A class of models proposed to explain the cosmic acceleration invoques
a scalar field playing the role of the dark energy, minimally or
non-minimally coupled to dark matter or gravity (see
e.g. \cite{de_models} and references therein). Despite of the variety
of the theoretical frameworks, all these models predict a time-varying
equation of state for the dark energy, usually represented through its
value at a given time, plus the first derivative in redshift or
cosmological scale factor \cite{linder,polarski}.

The dark energy equation of state is being constrained by a number of
experiments, such as Type Ia supernovae (SNIa) which have been the
first evidence for cosmic acceleration \cite{sn1a}, as well as Cosmic
Microwave Background (CMB) \cite{wmap} and Large Scale Structure (LSS)
\cite{tegmark}. The CMB total intensity power spectrum in particular
is sensitive to the dark energy equation of state mainly through a
projection effect induced by the variation of the distance to the last
scattering surface \cite{bacci}; since the latter is a redshift
integral between the present and last scattering, it washes out any
sensitivity to the time dependence of the dark energy equation of
state, probing just its redshift average.

A crucial benchmark in the era of precision cosmology is the measure
of the redshift behavior of the dark energy equation of state, fixing
the cosmological observables which are able to probe it at the epoch
of equivalence with matter or earlier, where most models predict
significant differences. A number of future probes aim at this goal,
including the Planck satellite ({\tt http://www.rssd.esa.int/PLANCK}),
the planned CMB polarization mission, as well as observation of SNIa
from space \cite{perlsnap} jointly with weak lensing surveys (see
e.g. \cite{refregier04} and references therein). The latter effect in
particular, i.e. the weak lensing induced on the background light by
forming structures is one of the most promising probes to investigate
the whole dark cosmological component, matter and energy, and is
gathering an increasing theoretical and observational interest (see
e.g. \cite{wl_review}); in particular, the weak lensing is potentially
relevant to investigate the high redshift behavior of the dark energy,
since the onset of cosmic acceleration overlaps in time with structure
formation.

In this work we consider the effect of the weak lensing on CMB
anisotropies. The theory of this process has been casted in the
context of the linear cosmological perturbation theory \cite{hu}, and
recently extended to cosmologies involving a scalar field dark energy
component with generalized kinetic energy and coupling to the Ricci
scalar \cite{acquaviva}. In particular, we concentrate on the
non-Gaussian distortion induced through the lensing effect on the CMB
total intensity anisotropies. An attempt to detect this signal
relating the data from the Wilkinson Microwave Anisotropy Probe (WMAP)
and the galaxy distribution in the Sloan Digital Sky Survey (SDSS) has
been done recently without success \cite{hirata}. This effect is
usually described in terms of the power injection in the CMB third
order statistics, represented in the harmonic domain by the CMB
bispectrum (see \cite{spergel_goldberg,komatsu} and references
therein); its relevance in constraining the redshift average of the
dark energy equation of state has been studied in detail
\cite{verde}. In our previous work \cite{giovi} we pointed out that
the bispectrum weak lensing signal is a promising probe of the high
redshift behavior of the dark energy, independently on the present
regime of acceleration. We studied the redshift distribution of the
weak lensing power injection in the CMB bispectrum in equilateral
configuration, demonstrating that it is vanishing at zero and
infinity, being relevant at intermediate redshift only, when cosmic
acceleration takes place. In this work we study the structure of the
whole bispectrum power from weak lensing, and set up a maximum
likelihood analysis fixing all cosmological parameters except those
related to the dark energy, to show how the bispectrum enhanced
sensitivity to the high redshift behavior of the expansion rate can be
used to break the projection degeneracy affecting the CMB power
spectrum.

The paper is structured as follows: in Sec. \ref{recall} we set our
framework by defining the dark energy parameterization and recalling
the basic aspects of the calculation of the CMB bispectrum signal from
weak lensing; in Sec. \ref{ratio} we analyze the tri-dimensional
multipole distribution of the bispectrum signal; in Sec. \ref{like} we
perform a likelihood analysis on simulated power spectrum and
bispectrum CMB data; in Sec. \ref{conclusions} we make our concluding
remarks.

\section{DARK ENERGY, WEAK LENSING AND CMB BISPECTRUM}
\label{recall}
It is now well understood that a dark energy component would mainly
affect the total intensity power spectrum of the CMB anisotropies
through a projection effect on the acoustic peaks
\cite{bacci}. Indeed, the dark energy dynamics may potentially alter
the distance to the last scattering surface, which provides a way to
estimate the dark energy content and its equation of state from CMB
measurements if a suitable dark energy model is given. In a different
approach, one could instead use CMB observations as a tool to infer
the underlying dark energy properties, without assuming its dynamics
following the prediction of a specific model. In this case
\cite{linder,polarski}, it is convenient to parameterize the evolution
of the dark energy equation of state as 
\begin{equation}
  \label{wz}
  w(z) = w_0 + (w_\infty - w_0) \frac{z}{1+z}\ ,
\end{equation}
where $w_0$ and $w_\infty$ are, respectively, the present and the
asymptotic ($z \rightarrow \infty$) values of the dark energy equation
of state. This is not the only parameterization, see for instance
\cite{corasaniti03}, but it is enough for our purposes here. Note that
the difference $(w_\infty - w_0)$ corresponds to the parameter $w_a$
in the original notation \cite{linder}. For reference, in
\cite{giovi}, we analyzed two scalar field dark energy models which,
at the level of background evolution, are very well reproduced by
appropriate choices of $w_{0}$ and $w_{\infty}$. The parameterization
(\ref{wz}) misses the fluctuations in the dark energy scalar field,
although that is not necessary here as we address the dependence of
the CMB three point statistics on the background expansion rate. The
comoving distance to the last scattering surface can be written as
\begin{equation}
  \label{e:dist}
  r(z_{lss}) = \frac{c}{H_0} \int_0^{z_{lss}}\frac{dz}{\sqrt{\Omega_m
      (1+z)^3 + \Omega_V e^{f(z)}}}\ ,
\end{equation}
where
\begin{equation}\label{e:fz}
  f(z) = 3 \int_0^z dz'\frac{1+w(z')}{1+z'}\ .
\end{equation}
In Eq. (\ref{e:dist}) we have restricted our analysis to a flat
universe and neglected radiation; $c$ is the speed of light, $H_0$
denotes the Hubble constant, $z_{lss}$ is the redshift of last
scattering surface and $\Omega_m=\Omega_b+\Omega_{cdm}$ is the matter
density today ($\Omega_b$ and $\Omega_{cdm}$ are respectively the
baryon energy density and the cold dark matter energy density). The
dark energy density is simply $\Omega_V=1-\Omega_m$.

However, reconstructing the behavior of $w(z)$ through the shift
induced on the CMB acoustic peaks turns out to be a complicate task:
as pointed out in \cite{verde}, the analysis of the power spectrum
alone is affected by a degeneracy between the dark energy equation of
state and $\Omega_V$. Such degeneracy becomes worse when the dark
energy dynamics is taken into account; typically, in the latter case
at least another parameter is introduced, as in Eq. (\ref{wz}): even
assuming $\Omega_V$ perfectly known, the two redshift integrals
prevent from measuring $w_0$ and $w_\infty$ separately from the CMB
power spectrum only. The use of higher order statistics \cite{giovi}
or a cross correlation between several observables (see
e.g. \cite{corasaniti04}) is needed to break this degeneracy.

Under the assumption of initial Gaussian conditions, the CMB
non-Gaussian diffuse signal is mainly generated by the
cross-correlation between the gravitational lensing and the Integrated
Sachs-Wolfe (ISW) effect. Such correlation causes a non-vanishing
power in the three point statistics, which may be described in terms
of the CMB bispectrum.

The three product of the harmonic expansion coefficients of the CMB
anisotropies in total intensity may be ensemble averaged to get the
bispectrum power on the angular scales specified by a triplet of
multipoles; the projected lensing potential allows to write down the
bispectrum dependence on the power spectrum of scalar perturbations
(see e.g. \cite{verde}); the result is
\begin{eqnarray}
  \label{e:genbis}
  \nonumber
  B_{l_1 l_2 l_3} = \sqrt{\frac{\left( 2l_1 + 1 \right) \left(2l_2 + 1
      \right) \left( 2l_3 + 1 \right)}{4\pi }} \left(
  \begin{array}{ccc}
    l_1 & l_2 & l_3 \\
    0 & 0 & 0
  \end{array} \right) \cdot \\
  \cdot \frac{l_1 \left( l_1 + 1 \right) - l_2 \left( l_2 + 1 \right)
    + l_3 \left( l_3 + 1 \right)}{2} C_{l_1}^{lss} Q \left(l_3 \right)
  + 5P\ ,
\end{eqnarray}
where the parenthesis denote Wigner 3J symbols, $C_{l_1}^{lss}$ is the
primordial power spectrum of total intensity anisotropies, and $5P$
indicates five permutation over the three multipoles. $Q(l)$ is
produced by the cross correlation between the gravitational lensing
and the ISW signal, expanded in spherical harmonics
\cite{spergel_goldberg,verde}:
\begin{equation}
  \label{e:ql}
  Q(l) \simeq 2 \int_0^{z_{lss}} dz \frac{r \left( z_{lss} \right) -
    r(z)}{r \left( z_{lss} \right)r^3 (z)} \left[ \frac{ \partial
      P_{\Psi} (k,z)}{\partial z} \right]_{k=\frac{l+1/2}{r(z)}}\ .
\end{equation}
Above the gravitational potential power spectrum is given by $P_{\Psi}
(k,z)=\left( \frac{3}{2} \Omega_m \right)^2 (H_0/ck)^4 P(k,z) (1+z)^2$
where $P(k,z)$ is the density power spectrum. We evaluate the
contribution of non-linearity to the matter power spectrum using the
semi-analytical approach of \cite{ma}; for a complete derivation of
Eq. (\ref{e:genbis}) and Eq. (\ref{e:ql}) see \cite{giovi,verde} and
references therein.

In writing Eq. (\ref{e:ql}), we used the Limber approximation; that is
widely used in the literature (see \cite{afs} and references therein)
and we perform here a numerical check of its validity. In the
calculations leading to (\ref{e:genbis}), one faces the product of
spherical Bessel functions. Those are suitably replaced by the
relation
\begin{equation}
  \label{e:besselj}
  j_l(x) = \sqrt{\frac{\pi}{2l+1}} \left[ \delta \left( l +
    \frac{1}{2} - x \right) + \Delta_l(x) \right]\ , 
\end{equation}
where $\delta$ is the Dirac's delta and the function $\Delta_l(x)$,
represents the difference between the Bessel function and the Dirac
delta itself; when integrating (\ref{e:besselj}) with regular
functions, one finds that the contribution from $\Delta_{l}$ vanishes
at high $l$s as $l^{-2}$. The Limber approximation consists in
neglecting the contributions of the $\Delta_l$ terms, which allows to
get to (\ref{e:ql}). More precisely, by substituting (\ref{e:besselj})
into the expression of the lensing-ISW correlation, one gets
\begin{eqnarray}
  Q_{true}(l) & = & Q(l) + (2l+1) \left\{ \int \frac{dr}{r^3}
  \frac{r_{lss} - r}{r_{lss} r} \int dr_1 \Delta_l(x)|_{x = ( l +
    \frac{1}{2} ) \frac{r_1}{r}} \left[ \Psi(k,r) \frac{\partial
      \Psi(k,r_1)}{\partial r_1} \right]_{k = \frac{l+1/2}{r}} +
  \right. \nonumber \\
  & + & \left. \int dr \frac{r_{lss} - r}{r_{lss} r} \int
  \frac{dr_1}{r_1^3} \Delta_l(x)|_{x = ( l + \frac{1}{2} )
    \frac{r}{r_1}} \left[ \Psi(k,r) \frac{\partial \Psi(k,r_1)}{\partial
      r_1} \right]_{k = \frac{l+1/2}{r_1}} \right\} +
  \mathcal{O}(l^{-3})\ ,
\label{Qtrue}
\end{eqnarray}
where $\Psi(k,r)$ is the cosmological gravitational potential, defined
as the metric fluctuations in absence of anisotropic stress; the
underlying assumption of Gaussian distribution of cosmological
perturbations eliminates one of the two integrals in the Fourier
wavenumbers. The last term, representing the error on $Q(l)$, presents
an overall behavior as $l^{-1}$, coming from the combination of
$\Delta_{l}$ and the $2l+1$ term in front of the integral. We
quantified the error induced by the Limber approximation by computing
$Q(l)$ with and without that, considering a $\Lambda$CDM cosmological
model consistent with the WMAP first year data \cite{wmap}. We find an
error which is about $1\%$ at $l=10$, falling down approximately as
$1/l$ afterwards as predicted by our expectation (\ref{Qtrue}); we
stress that $l=10$ is the lowest multipole considered in our analysis,
as it was done before (see \cite{verde}) in order not to approach too
low multipoles where the Limber approximation may not be
satisfactory. Our results are also consistent with \cite{afs}, quoting
a $2-3\%$ error at $l=3$, corresponding to about $1\%$ at $l=10$ with
an $1/l$ scaling, as we find. In conclusion, we verify that the Limber
approximation holds at percent or better in our analysis, and we
exploit that in the rest of the work.

\section{THE STRUCTURE OF THE TRI-DIMENSIONAL BISPECTRUM}
\label{ratio}
In this Section, we investigate the physical properties and
phenomenology of the weak lensing bispectrum signal in its most
general configuration.

In \cite{giovi}, only ``equilateral'' geometries were considered
(i.e., $l_1=l_2=l_3$ in Eq. (\ref{e:genbis})): however, as we discuss
in the following, it is greatly restrictive to consider that
configuration only, expecially from the point of view of the signal to
noise ratio. As discussed in detail in \cite{hu}, the cosmic variance
and instrumental noise act on the $B_{l_1 l_2 l_3}$ elements
essentially as a three product of the sum $C_l^{lss}+C_l^{noise}$,
where the last term represents the instrumental noise
contribution. Therefore, the bispectrum signal to noise ratio is
defined as
\begin{equation}
  \left( \frac{S}{N} \right)^{2} = \sum_{l_1,l_2,l_3}
  \frac{B_{l_1 l_2 l_3}^2}{n_{l_1 l_2 l_3} C_{l_1} C_{l_2} C_{l_3}}\ ,
  \label{e:snratio}
\end{equation}
where the constant $n_{l_1 l_2 l_3}$ is $6$ for equilateral
configurations, $2$ if two multipoles are equal (isosceles triangles)
and $1$ if all multipoles are different (scalene triangles). Note
that, in order to make use of the small angle approximation
\cite{verde}, we have to adopt a lower limit on $l$, indicated as
$l_{min}$; as in \cite{giovi} we fix $l_{min}=10$ in the
following. The results do not depend at all on this choice. From
Eq. (\ref{e:snratio}) we see that in the equilateral configuration we
are summing $l_{max}$ terms, while in the most general case the signal
to noise ratio receives contributions from about $l_{max}^3$ terms:
for this reason, it is clear that the ``scalene'' configuration
contains much more information, just because the sum (\ref{e:snratio})
is dominated by those coefficients.

As it is evident in equation (\ref{e:genbis}), the weak lensing
relates the primordial signal $C_l^{lss}$ and the lensing kernel given
in (\ref{e:ql}) on different scales. The first exhibits an
oscillatory, decaying behavior around $l^{-2}$, at least for $l
\lesssim 1000$ i.e. before the damping tail; the second is a straight
power law decaying faster, followed by a change of sign and a
shallower decay occurring on the scale of a few hundreds in $l$ as a
result of the domination of the non-linear tail in the density power
spectrum \cite{giovi}. The combination of the two behaviors, as a
general function of the three multipoles, distributes the oscillatory
power of $C_l^{lss}$ on scales different from the ones of the
primordial acoustic oscillations, creating a complex pattern of peaks
and valleys which is the recurring feature of the weak lensing
bispectrum signal. Such structure is not distributed over the whole
$(l_1,l_2,l_3)$ space, since geometrical constraints greatly reduce
the avaliable domain. A physical constraint making the signal
vanishing is represented by the transition from linear to non-linear
power domination in (\ref{e:ql}).

\begin{figure}
  \centering
  \includegraphics{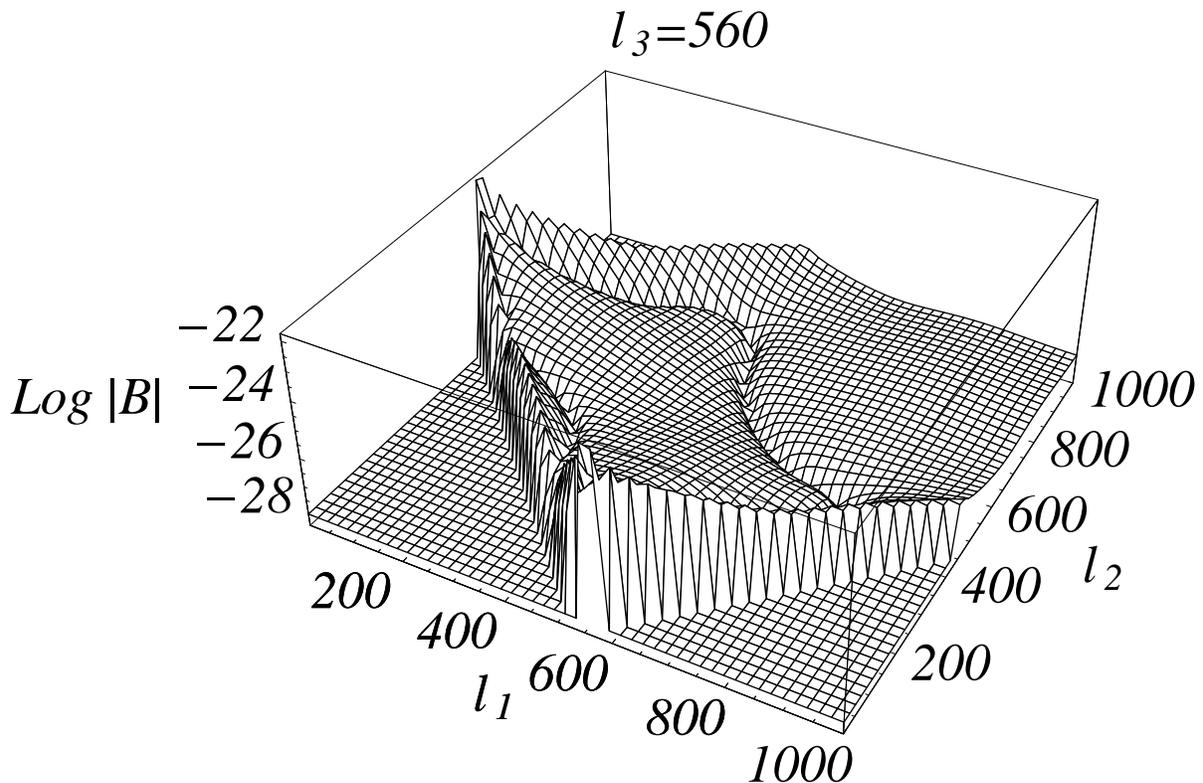}
  \caption{Example of distribution of tri-dimensional bispectrum
    coefficients. The domain is determined by geometrical constraints
    in the Wigner symbols, while the ``canyon" is due to the
    transition between linear and non-linear power in the density
    power spectrum. The primordial acoustic oscillations are projected
    on multiple angular scales by the weak lensing effect.}
  \label{f:bis_3d}
\end{figure}

It is convenient to give an illustration of that phenomenology
providing an example of tri-dimensional view of the bispectrum signal;
in Fig. \ref{f:bis_3d} we plot the absolute value of the bispectrum
coefficients for a cosmological model with $\Omega_V=0.73$, $w_0=-1$
and $w_\infty=-0.8$; the remaining cosmological parameters are set
accordingly to the cosmic concordance model, $H_0=72$ km/s/Mpc,
$\Omega_b=0.046$, $\Omega_{cdm}=1-\Omega_V-\Omega_b$, reionization
optical depth $\tau=0.11$, scalar perturbations only with spectral
index $n_s=0.96$. The signal is dimensionless ($Q(l)$ is a number and
$C_l^{lss} \propto (\delta T/T)^2$) and valid for a cosmology with
$\Omega_V=0.73$, $w_0=-1$ and $w_\infty=-0.8$, corresponding to our
fiducial model as we explain later. The first thing to note is that
the bispectrum is vanishing for multipoles triplets which do not
satisfy the triangular relation (Wigner 3J symbols are zero) as well
as for configurations where the sum of multipoles is odd. In the case
shown, $Q\left(l_3\right)=0$; a variation of $l_3$ changes the regions
where the triangularity relation is not verified: if $l_3$ is
increased, the bispectrum domain is squeezed along the direction
$l_1=l_2$ and is stretched in the orthogonal direction; if $l_3$ is
decreased, we obtain the opposite behavior.

The two main features discussed above are well evident. First, the
sort of ``canyon'' representing the fingerprint of the transition
between the linear and non-linear regime in the power spectrum
derivative in (\ref{e:ql}), yielding a change of sign appearing
``cuspidal'' in module; a ``valley'' appears close and aligned with
the left border of the bispectrum domain, this time determined
entirely by the behavior of the signal which is rising at the lowest
multipoles \cite{giovi}. Second, the re-projection of the acoustic
oscillations due to the mixing (\ref{e:genbis}): the whole power in
the central region of the figure, roughly between the valley and the
canyon, comes mainly from the first acoustic oscillation in the power
spectrum, which gets projected on a large domain of multipole
triplets, i.e. angular scales. The phenomenology is therefore markedly
richer with respect to the case of the ordinary power spectrum. On the
other hand, the weak lensing signal is a second order effect in terms
of cosmological perturbations, and the cosmic variance alone does not
allow the detection of the single bispectrum
coefficient. Nevertheless, a satisfactory signal to noise ratio is
achieved by summing over all the different configurations.

\begin{figure}
  \centering
  \includegraphics[clip=true]{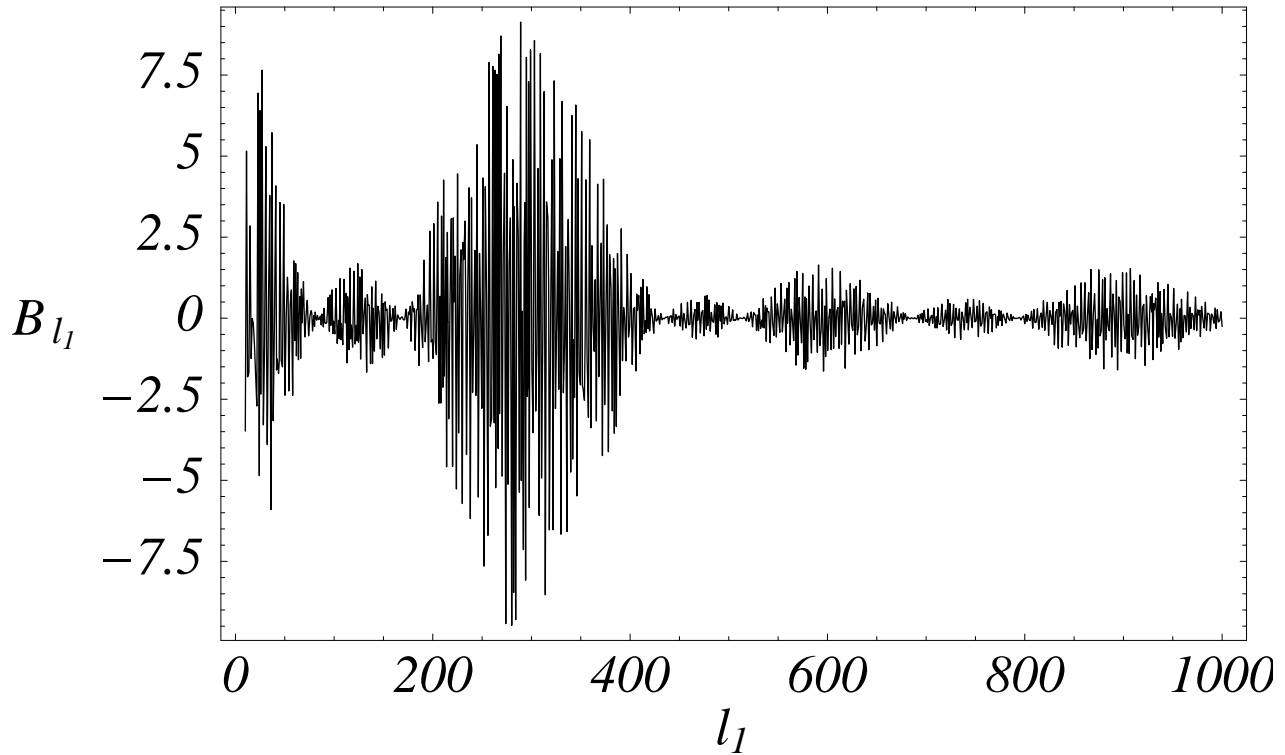}
  \caption{Bispectrum in units of $10^{-20}$ as function of the
    multipole $l_1$ only, obtained by marginalization over $l_2$ and
    $l_3$. The small scale, negative and positive oscillations are due
    to the Wigner 3J symbols. The large scale modulation is instead a
    physical feature related to the primordial acoustic peaks.} 
  \label{f:l1-bis}
\end{figure}

The tri-dimensional information may be compressed in several ways; a
uni-dimensional quantity which contains the contributions from all the
possible configurations is
\begin{equation}
  \label{e:compressbis}
  B_{l_1}=\sum_{l_2,l_3=l_{min}}^{l_{max}} B_{l_1 l_2 l_3}\ ,
\end{equation}
where we take $l_{min}=10$ and $l_{max}=1000$. In Fig. \ref{f:l1-bis}
we plot $B_{l_1}$. The figure is representative of the angular scales
from which the signal takes most of the contribution, which are close
to the maximum primordial power in the first acoustic peak. This is
consistent with previous analysis \cite{spergel_goldberg}. The
remaining part of the signal is made of a relevant part at low
multipoles, probably due to the rise of the overall bispectrum power
(see Fig. \ref{f:bis_3d}), plus a series of oscillations with varying
amplitudes. The latter seem still related to the series of peaks in
the primordial power. Finally, the positive and negative high
frequency oscillations are entirely due to the behavior of the 3J
Wigner symbols.

\begin{figure}
  \centering
  \includegraphics{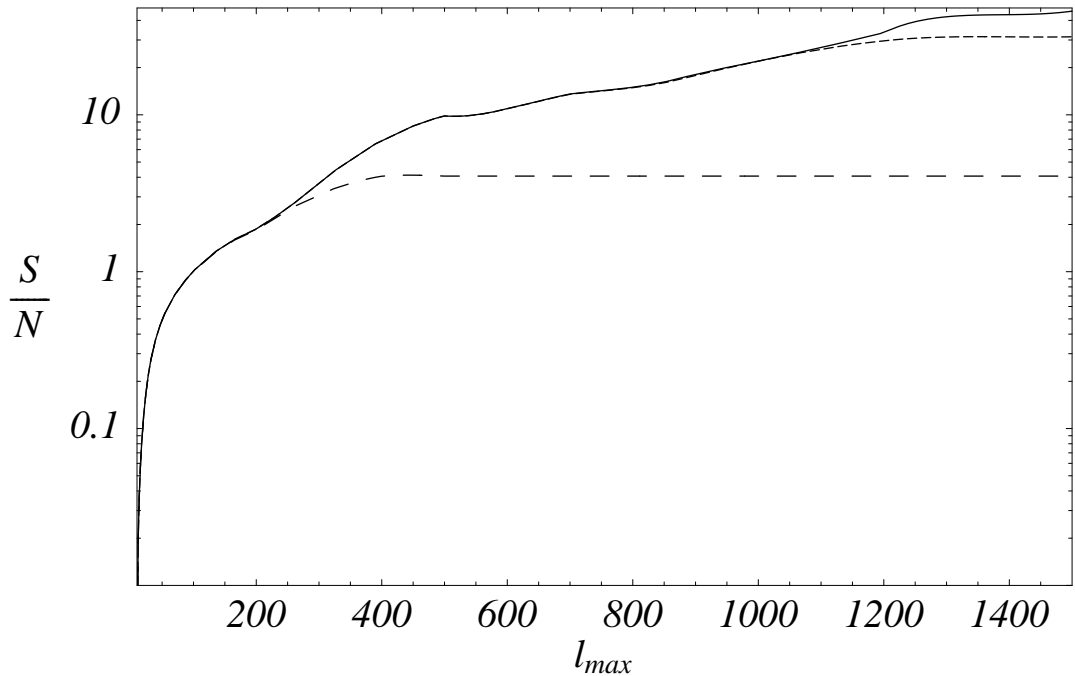}
  \caption{Bispectrum signal to noise ratio as function of $l_{max}$
    for a cosmic variance limited experiment (solid line), Planck two
    years nominal noise (short dashed line) and WMAP two years nominal
    noise (long dashed line). The cosmology corresponds to our
    fiducial model with $\Omega_V=0.73$, $w_0=-1$ and $w_\infty=-0.8$;
    the other parameters are fixed accordingly to the cosmic
    concordance model.}
  \label{f:snratio}
\end{figure}

The information described in Fig. \ref{f:bis_3d} and \ref{f:l1-bis} is
totally compressed when the signal to noise ratio, defined in
(\ref{e:snratio}), is computed. It is shown in Fig. \ref{f:snratio}
for three cases: two years WMAP nominal noise, two years Planck
nominal noise (as in Table 1 in \cite{balbi}) and cosmic variance
only; all the cases considered do not take into account systematics
and foregrounds, and an all sky coverage is assumed. The plateau means
that no extra information is added extending the sum (\ref{e:snratio})
at very high $l$s, where the signal vanishes below cosmic variance and
noise. WMAP is cosmic variance limited up to $l \simeq 300$, while
Planck up to $l \simeq 1000$. Of course this discussion does not take
into account foreground or systematics contamination, which in
practice may represent major obstacles against the actual
detection. In the following, we consider only the effect of cosmic
variance, and probe the bispectrum up to $l_{max}\simeq 1000$; the
results are therefore representative of an experiment with the nominal
performance of Planck.

\section{COMBINING POWER SPECTRUM AND BISPECTRUM: LIKELIHOOD ANALYSIS
  AND RESULTS}
\label{like}
In this Section, we discuss how the bispectrum data improve the CMB
sensitivity to the high redshift behavior of the dark energy.

As already mentioned, the power spectrum suffers the degeneracy of the
comoving distance to last scattering surface relatively to different
values of the dark energy parameters. One of the most important
aspects of the CMB bispectrum from weak lensing is the fact that it
correlates the primordial temperature anisotropies with later physical
processes, i.e. the formation of structures), probing the dark energy
behavior at that epoch \cite{giovi}.

Let us describe here our likelihood analysis. Most of the cosmological
parameters are fixed according to the cosmological concordance model;
we shall perform only a variation of the dark energy parameters
$\Omega_V$, $w_0$ and $w_\infty$, which allows us to build a grid of
$C_l$ and $Q(l)$ values. Then, spanning over our parameter space, we
calculate the bispectrum by mean of the relation
(\ref{e:genbis}). Once we have that as a function of $\Omega_V, w_0$
and $w_\infty$, we compute a three parameters likelihood both for the
power spectrum and for the bispectrum; the combined analysis is simply
obtained by multiplying the two of them.

The other cosmological parameters are set to their values for our
fiducial model, described earlier. We evaluate the likelihood as usual
as
\begin{equation}
  \like_{s,b} = A_{s,b} \exp \left( -\frac{\chi_{s,b}^2}{2} \right)\ ,
\end{equation}
where the subscripts refer, respectively, to the power spectrum $(s)$
and to the bispectrum $(b)$, $A_{s,b}$ is a normalization factor and
$\chi_{s,b}^2$ are functions of the dark energy parameters defined as
\begin{eqnarray}
  \chi_s^2 & = & \sum_{l=2}^{1000}\left[ \frac{C_l^t-C_l^f}{\sigma_l^s}
    \right]^2\ ,  \\
  \chi_b^2 & = & \sum_{l_1,l_2,l_3=10}^{1000} \left[ \frac{B_{l_1 l_2
	l_3}^t-B_{l_1 l_2 l_3}^f}{\sigma_{l_1 l_2 l_3}^b} \right]^2\ .
\end{eqnarray}
In the expression above, we describe spectrum and bispectrum as
Gaussian variables \cite{gangui,takada}: that corresponds to ignore
the possible non-Gaussianity arising in the early universe, as well as
to assume a Gaussian distribution for the bispectrum coefficients,
exploiting the argument which states that in the case of weak lensing
the signal is caused by many independent events. The superscript $t$
refers to the theoretical model, while $f$ refers to our fiducial
model. As we already mentioned, we limit our analysis to multipoles
smaller than 1000, being consistent with a Planck-like experiment with
nominal performance (see Fig. \ref{f:snratio}). For the power
spectrum, the well known expression of the cosmic variance is $\left(
\sigma_l^s \right)^2=2/(2l+1) C_l^2$; for the bispectrum, $\left(
\sigma_{l_1 l_2 l_3}^b \right)^2=n_{l_1 l_2 l_3} C_{l_1} C_{l_2}
C_{l_3}$. We also neglect the correlation between the two observables
as it is induced by higher order statistics \cite{takada}, which
allows us to write down the combined likelihood as
\begin{equation}
  \label{e:like}
  \like_c = \like_s \like_b = A_c\exp \left( -\frac{\chi_s^2 +
    \chi_b^2}{2} \right)\ ,
\end{equation}
where $A_c$ normalizes the combined likelihood.

We report the results of our likelihood analysis in three main cases,
with one, two and three free parameters. As we show later this
approach is necessary to understand how different the basins of
degeneracy of spectrum and bispectrum are, and how their combination
actually break the projection degeneracy affecting the power
spectrum. We use the following priors: $0 \le \Omega_V \le 1$, $-1 \le
w_0 \le -0.6$, $-1 \le w_\infty \le -0.6$. When $\Omega_V$ varies, the
matter abundance is changed to keep flatness and leaving $\Omega_b$
unchanged.

\subsection{One parameter likelihood}
This is the simpler case, where two of the three dark energy
parameters are fixed to their fiducial values, while the third one is
allowed to vary. In that case, regardless of which parameter varies,
the bispectrum doesn't improve the likelihood analysis of the power
spectrum alone. The reason is simply the reduced signal to noise ratio
of the bispectrum. When just one parameter vary, even $w_\infty$, the
tiny changes in the power spectrum signal dominate the exponential in
the combined likelihood; note that when $w_\infty$ varies, the
relative changes in the bispectrum are stronger than in the power
spectrum case \cite{giovi}. However, when the noise is taken into
account, such advantage is canceled and the joint likelihood is almost
identical to $\like_s$. As we shall see in the next sub-section, the
results change greatly when a multi-dimensional analysis is
performed. The reason is that the basins of degeneracy of spectrum and
bispectrum open in markedly different ways, making their combination
advantageous.

\subsection{Two parameters likelihood}
\label{two}
Here we analyze the likelihood behavior with respect to the variation
of two dark energy parameters.

\begin{figure}
  \centering
  \resizebox{\hsize}{!}{\includegraphics[clip=true]{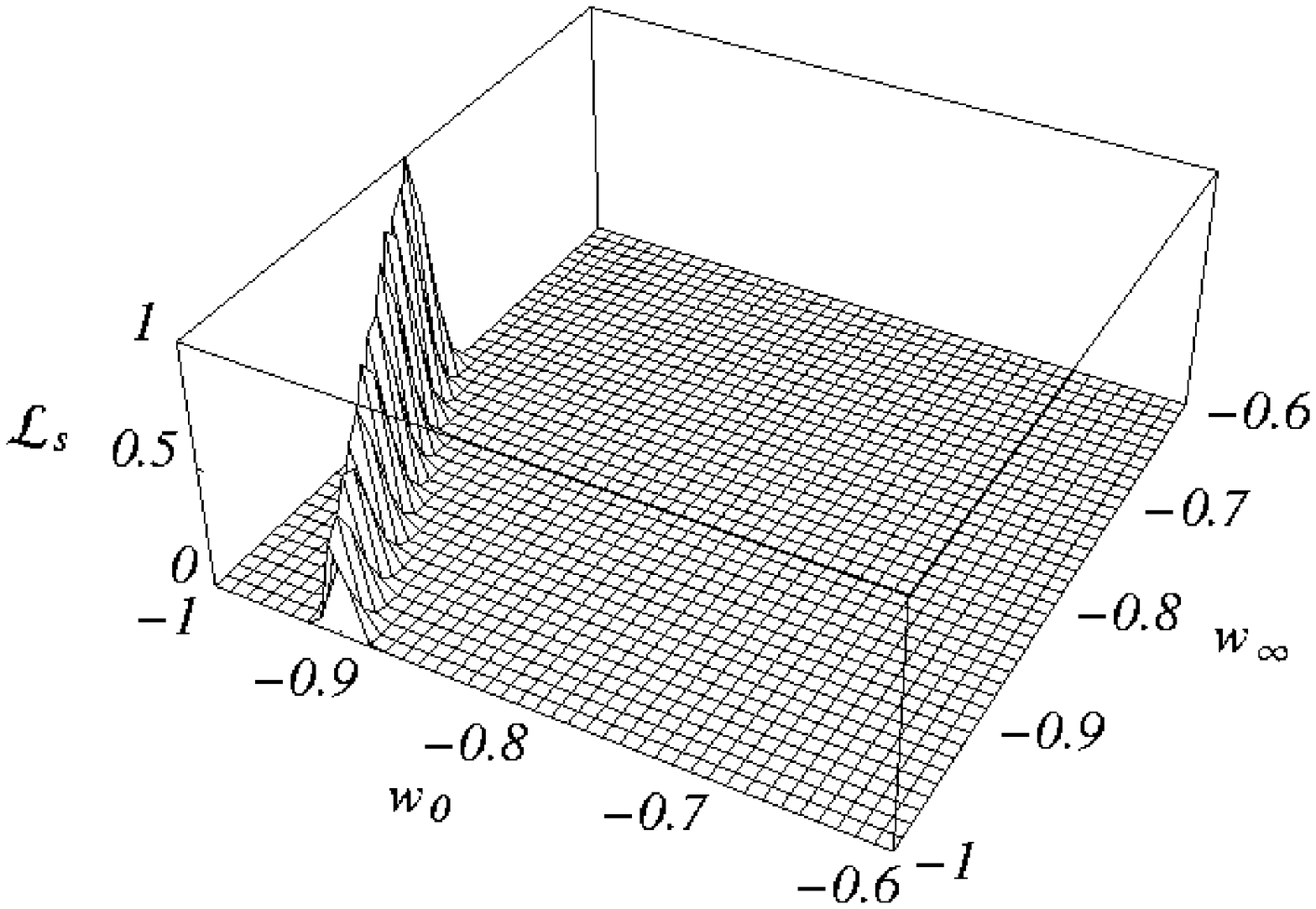}
    \includegraphics[clip=true]{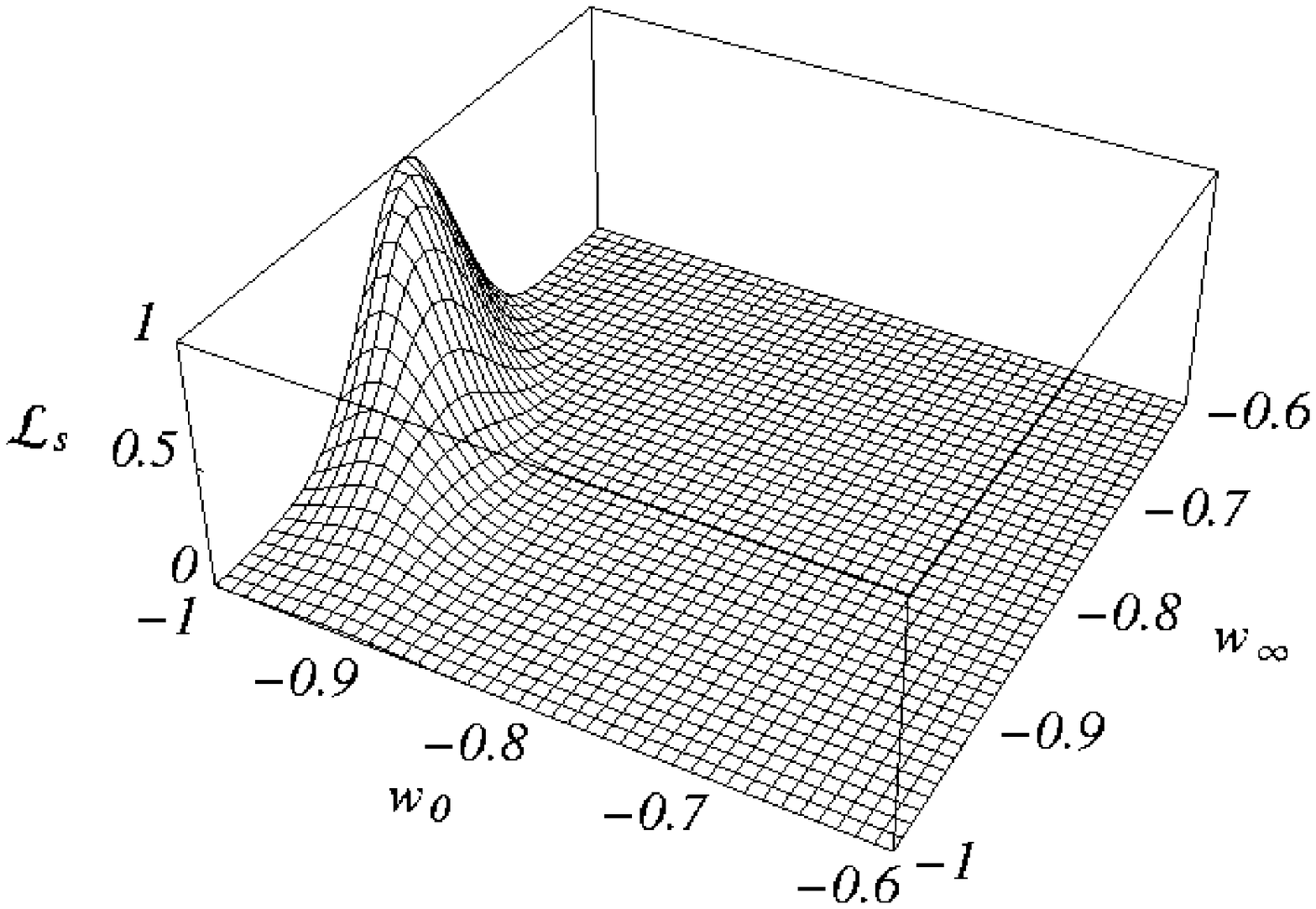}
    \includegraphics[clip=true]{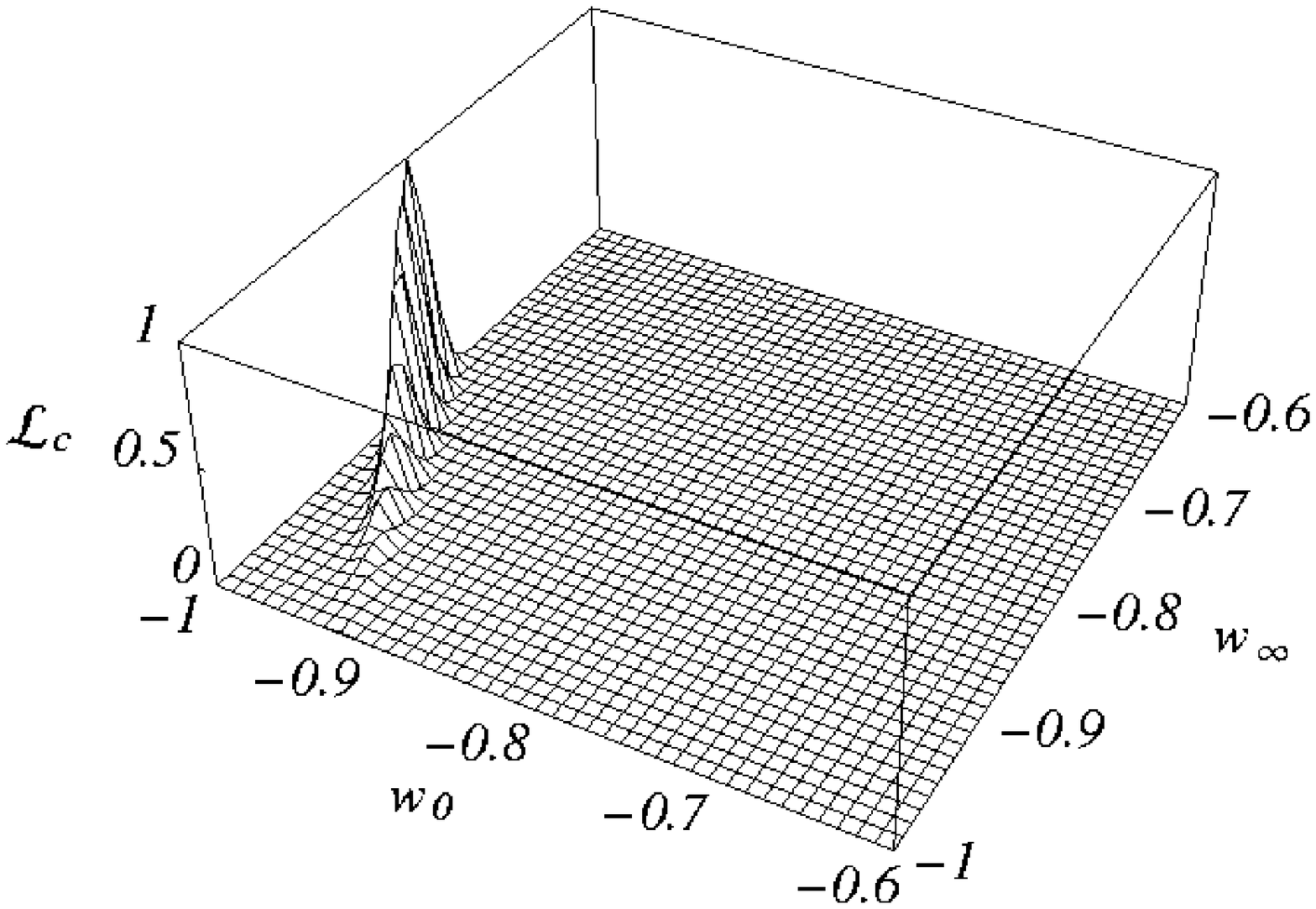}}
  \caption{Likelihoods as function of $w_0$ and $w_\infty$ with
    $\Omega_V=0.73$; from left to right: power spectrum only,
    bispectrum only and their combination.}
  \label{f:like2pwwi}
\end{figure}

\begin{figure}
  \centering
  \resizebox{\hsize}{!}{\includegraphics[clip=true]{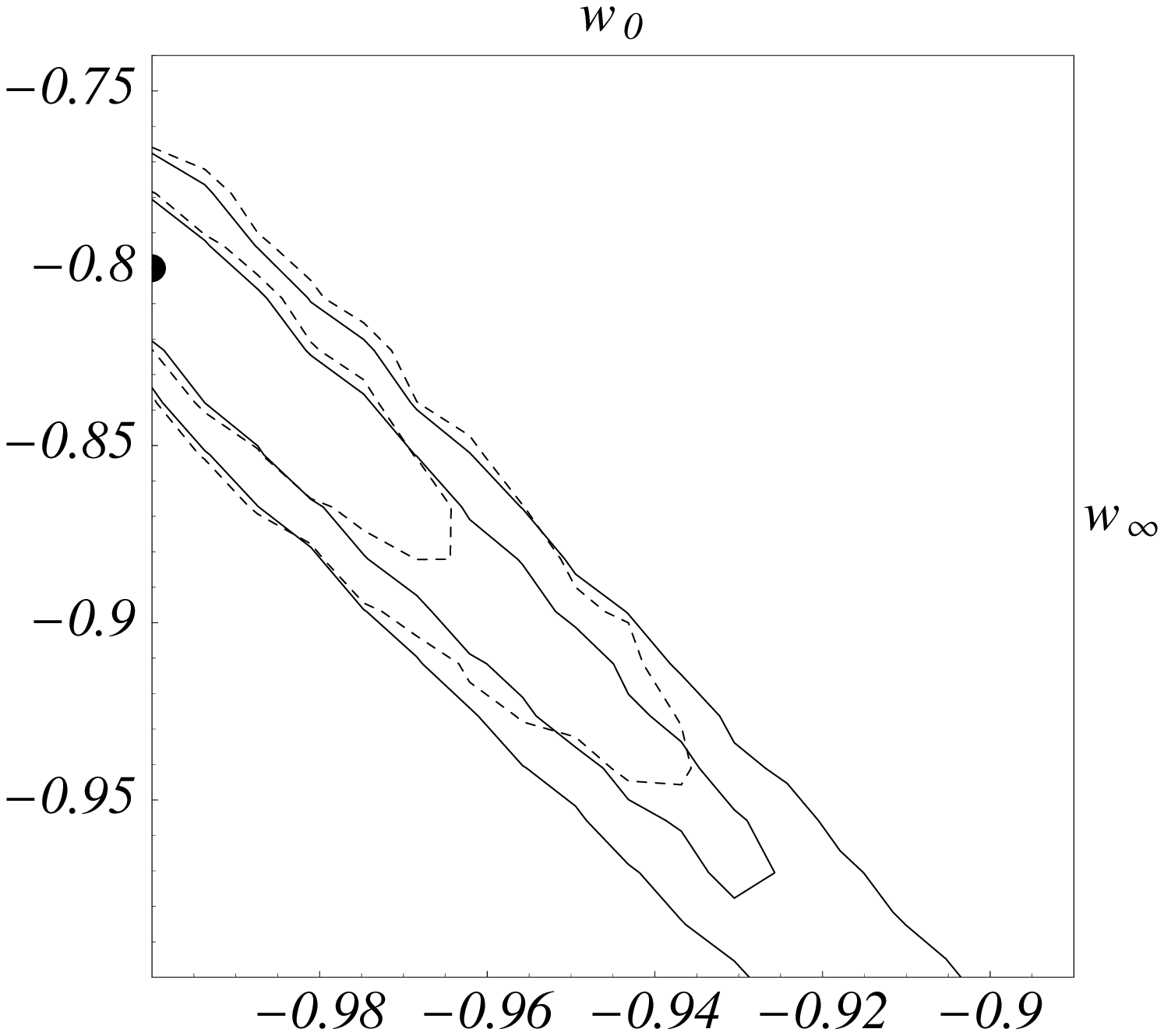}
    \includegraphics[clip=true]{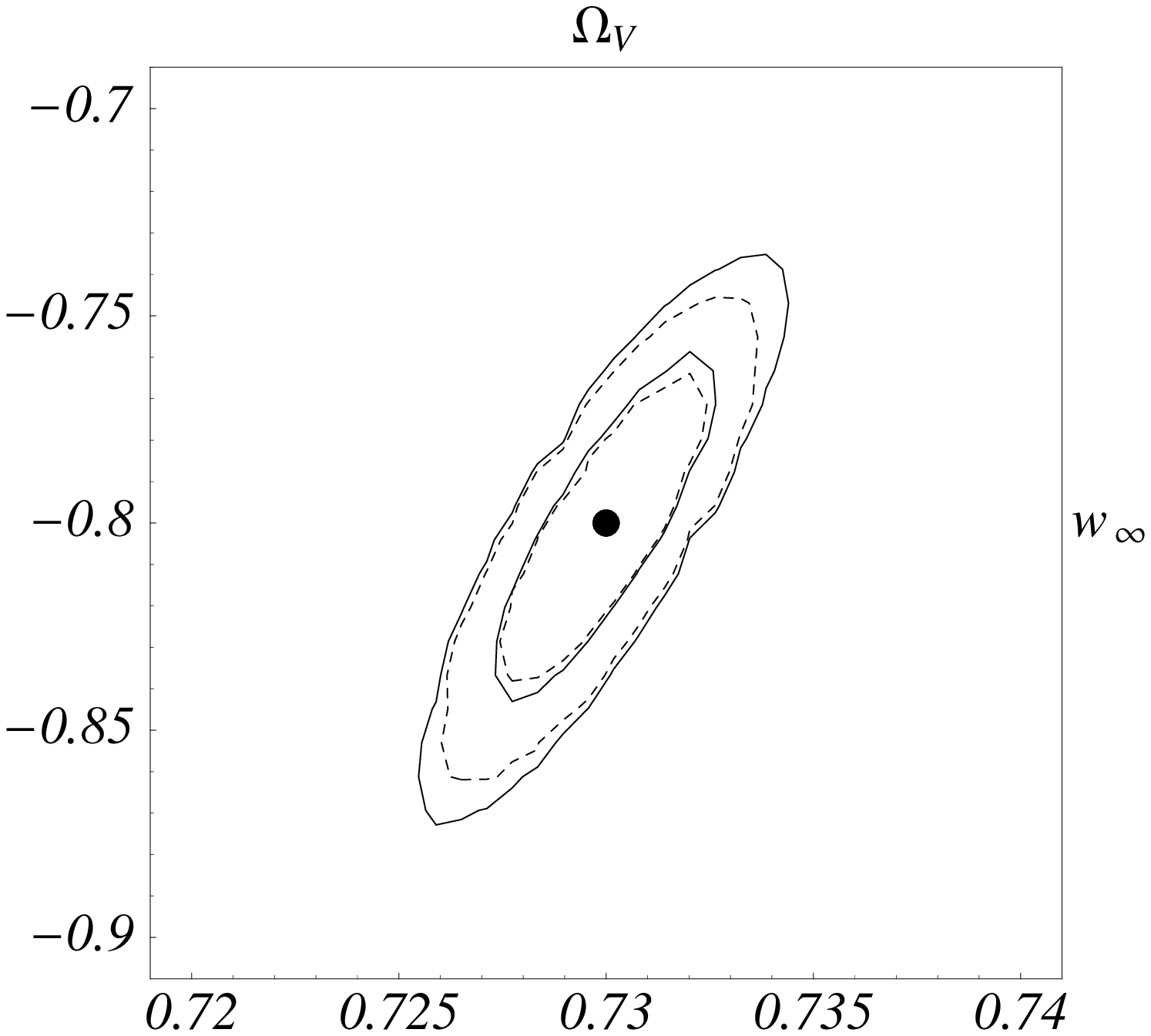}
    \includegraphics[clip=true]{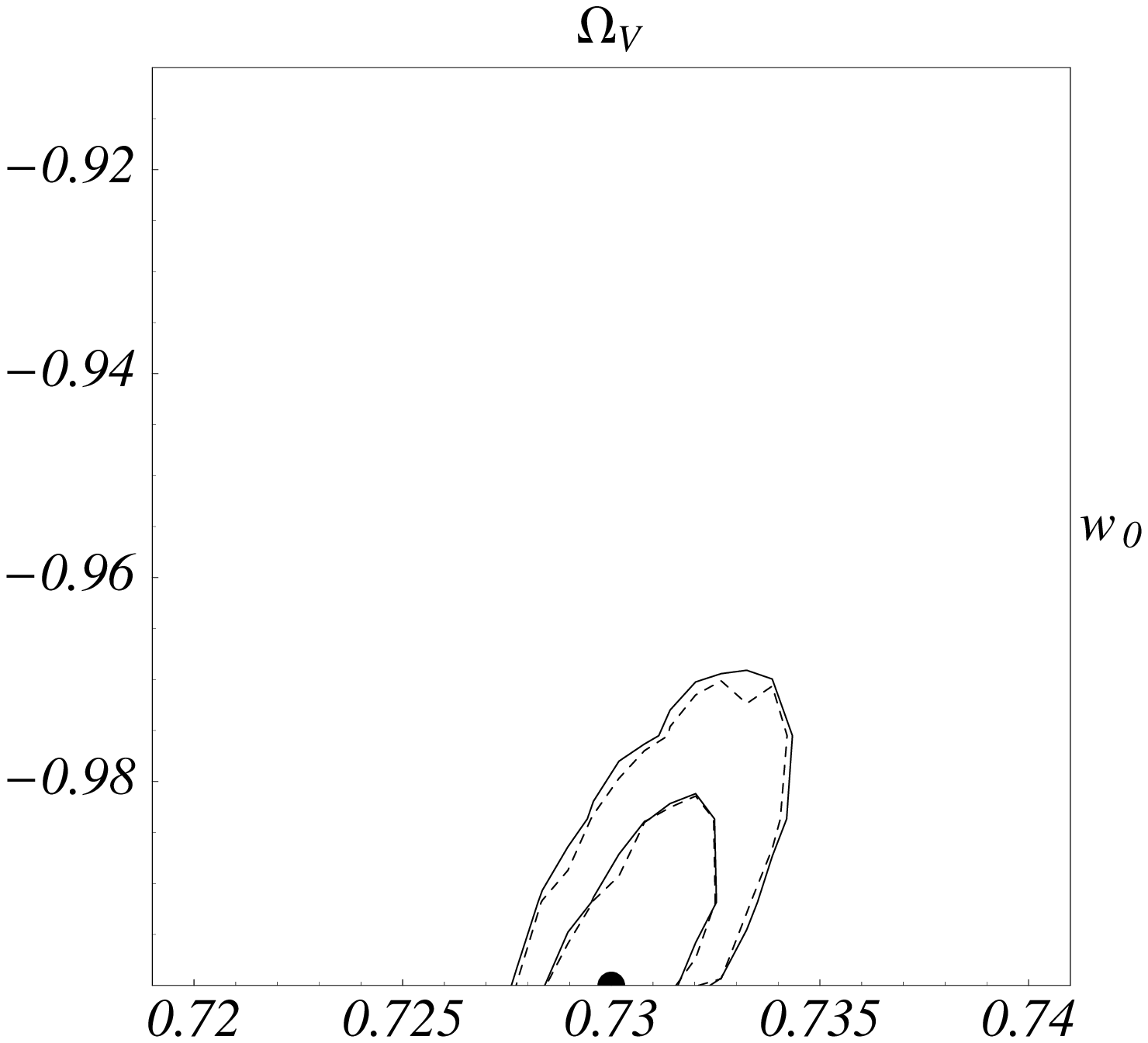}}
  \caption{Likelihood confidence levels at $1\sigma$ (innermost
    contours) and $2\sigma$ (outermost contours) for power spectrum
    only (solid line) and for power spectrum and bispectrum (dashed
    line). From left to right: marginalization over $\Omega_V$,
    marginalization over $w_0$, marginalization over $w_\infty$. The
    dot is the fiducial model and it is located at $(\Omega_V=0.73,
    w_0=-1, w_\infty=-0.8)$.}
  \label{f:like_2d}
\end{figure}

Let's begin by fixing $\Omega_V=0.73$; though the power spectrum
likelihood is very narrow (see Fig. \ref{f:like2pwwi} left panel), it
is degenerate along a specific direction, where different values of
$w_0$ and $w_\infty$ produce similar values of the function $e^{f(z)}$
in Eq. (\ref{e:dist}) at the epoch of the structure formation. For
this reason, the power spectrum is able to exclude a wide region of
the parameter space, but it is unable to constrain $w_0$ and
$w_\infty$ separately. The bispectrum likelihood is non-zero over a
wider region of the parameter space, but vanishes on the region where
the power spectrum is degenerate (Fig. \ref{f:like2pwwi} center
panel); while the projection degeneracy affecting the power spectrum
is well understood, it is more difficult to track the dark energy
variables in the bispectrum calculation, as they enter in several
aspects, i.e. distances and perturbation growth. The only clear
feature is the vanishing of the lensing power at present and infinity
\cite{giovi}, making the bispectrum sensitive to $w_\infty$ more than
to $w_0$, as in Fig. \ref{f:like2pwwi}; the likelihood shape, rather
elongated on the $w_0$ direction, and sharper on the $w_\infty$ one,
is clearly consistent with our expectation. As a result, the
degeneracy basins of spectrum and bispectrum are misaligned, allowing
the separate measure of $w_0$ and $w_\infty$ from their combination
(Fig. \ref{f:like2pwwi}, right panel). In the left panel of
Fig. \ref{f:like_2d} we show the contour plot of the power spectrum
and bispectrum likelihoods; as we can see, the gain is evident.

In the center panel and in the right one of Fig. \ref{f:like_2d} we
show the contour levels of the other two likelihood analysis, fixing
$w_0$ and $w_\infty$ respectively. In those cases, getting the
bispectrum into the analysis does not improve substantially the
constraints. The reason is that fixing one parameter in the
exponential $e^{f(z)}$ in Eq. (\ref{e:dist}) virtually break the power
spectrum projection degeneracy: more specifically, separate changes in
$w_0$ or $w_\infty$ cannot be compensated by changes in $\Omega_V$,
since the latter changes the matter abundance, inducing changes in the
power spectrum shape which again dominate the exponential in the joint
likelihood. As we shall see now, the latter reasoning does not apply
when all the three parameters vary, causing a relevant gain for all of
them in the combined analysis.

\subsection{Three parameters likelihood}
\begin{figure}
  \centering
  \resizebox{\hsize}{!}{\includegraphics[clip=true]{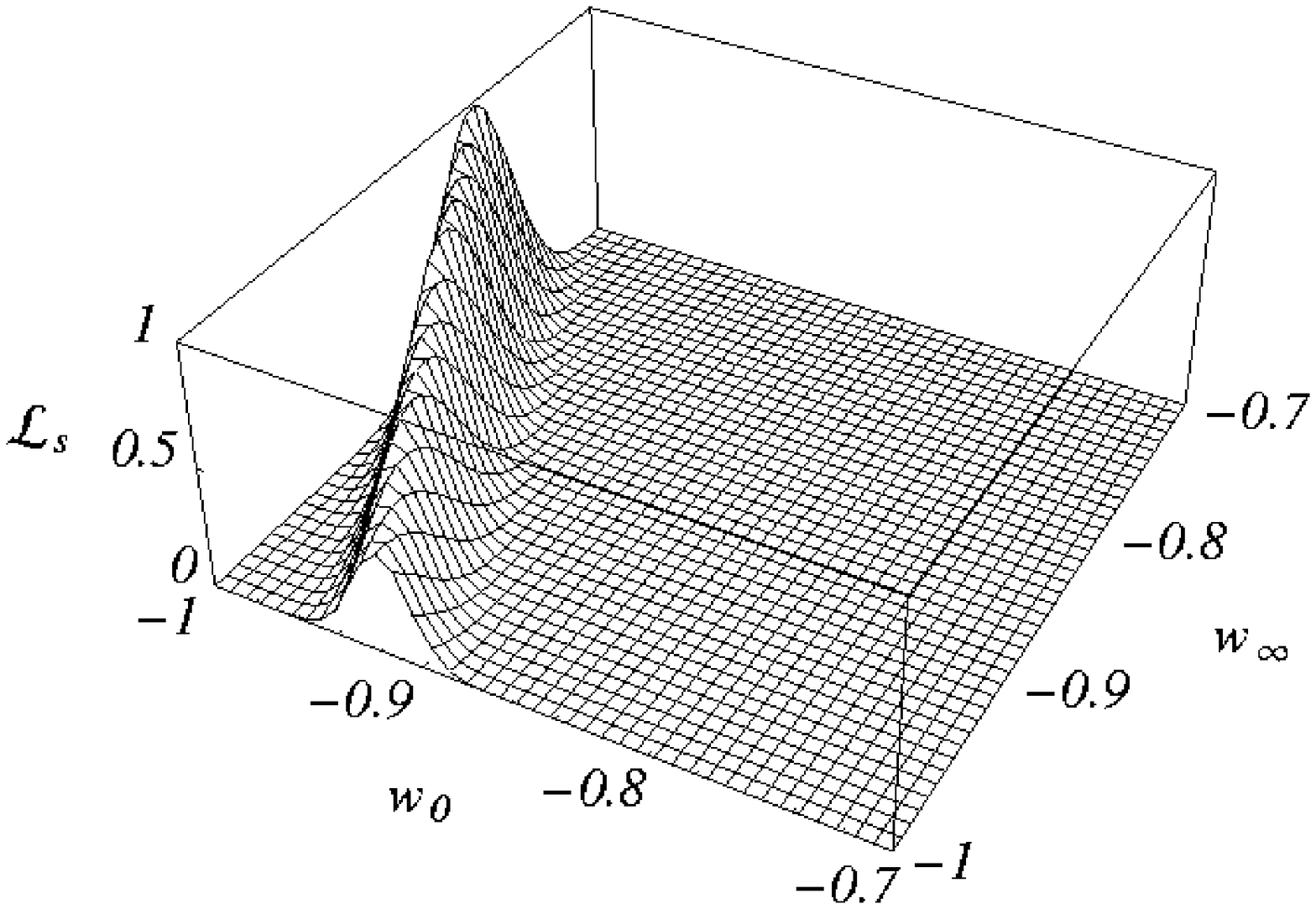}
    \includegraphics[clip=true]{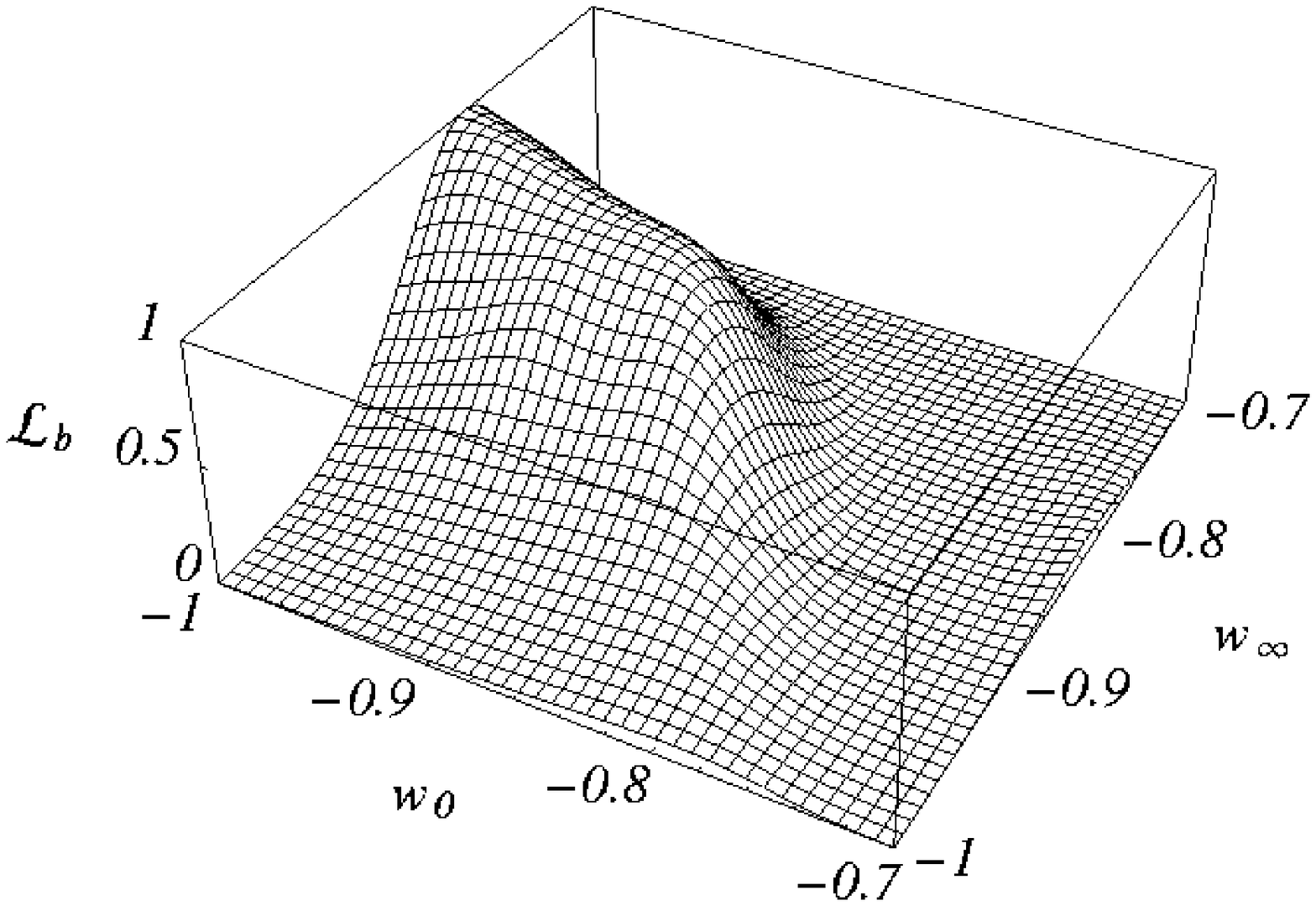}}
  \resizebox{\hsize}{!}{\includegraphics[clip=true]{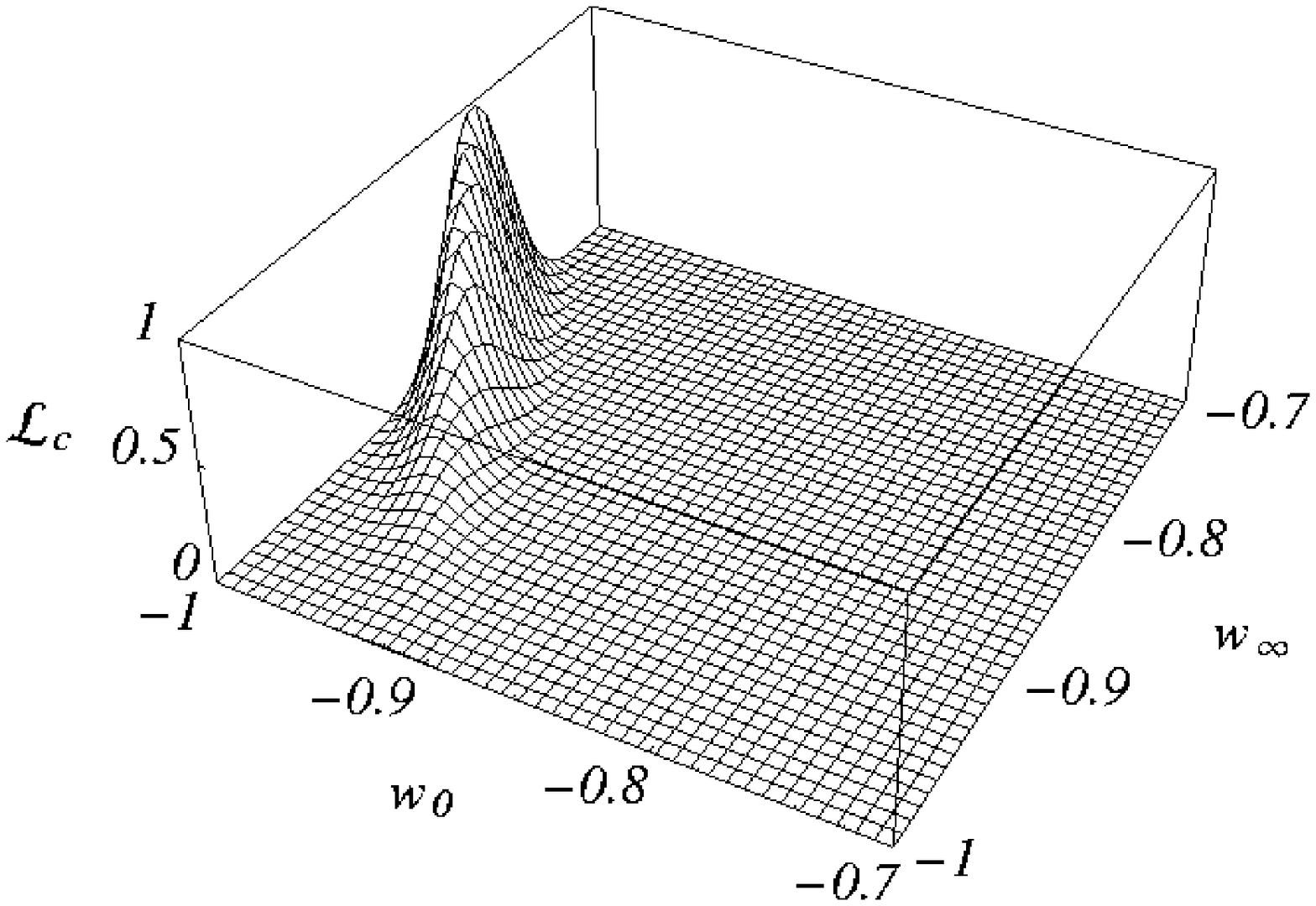} 
    \includegraphics[clip=true]{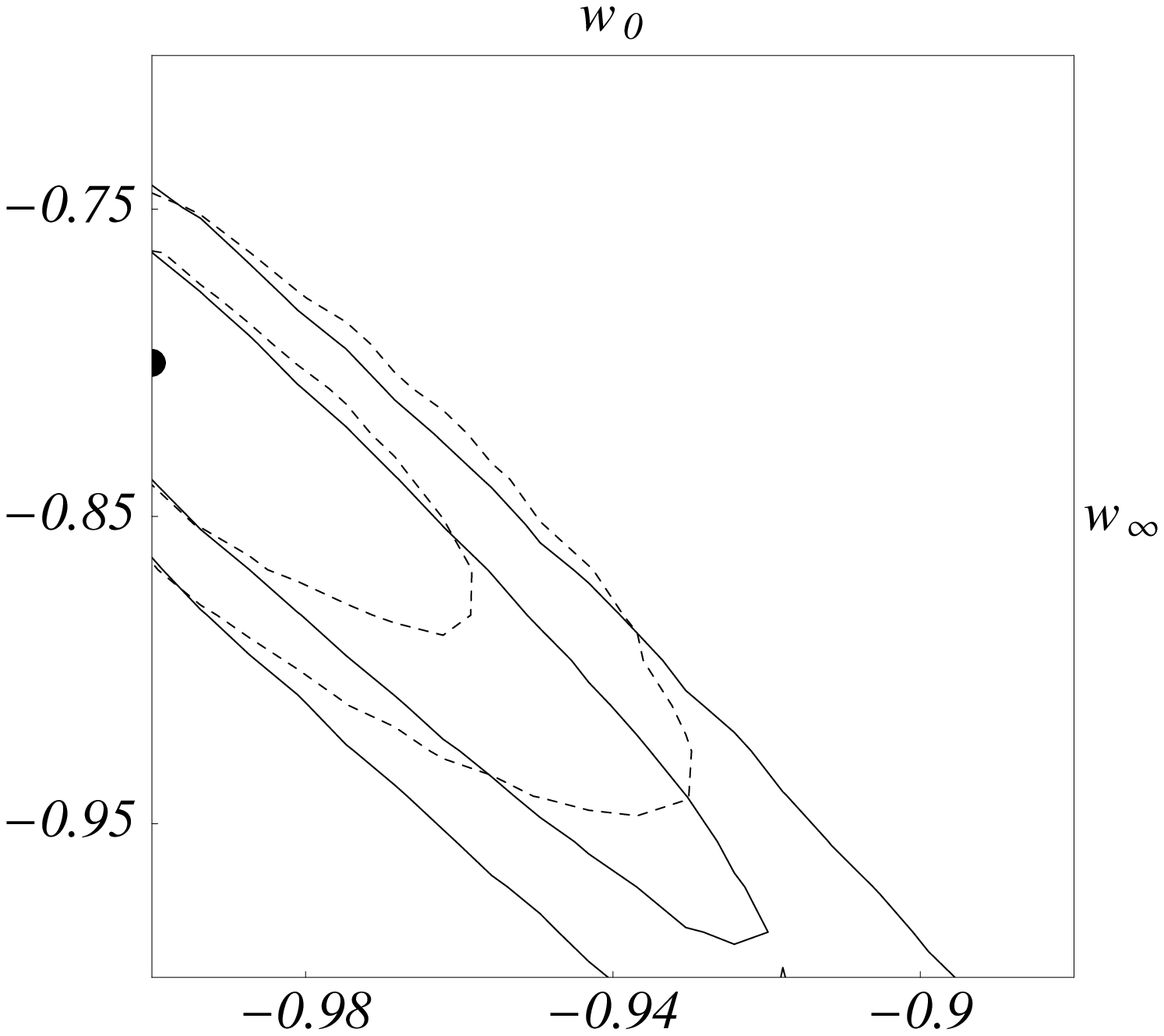}}
  \caption{Likelihood marginalized over $\Omega_V$ for the power
    spectrum only (top-left), bispectrum only (top-right) and both
    (bottom-left). In the bottom-right panel we plot the contour
    levels at $1\sigma$ (innermost contours) and $2\sigma$ (outermost
    contours) for the power spectrum by itself (solid line) and
    combined with the bispectrum (dashed line), the dot is our
    fiducial model.}
  \label{f:w0wi_3d}
\end{figure}

\begin{figure}
  \centering
  \resizebox{\hsize}{!}{\includegraphics[clip=true]{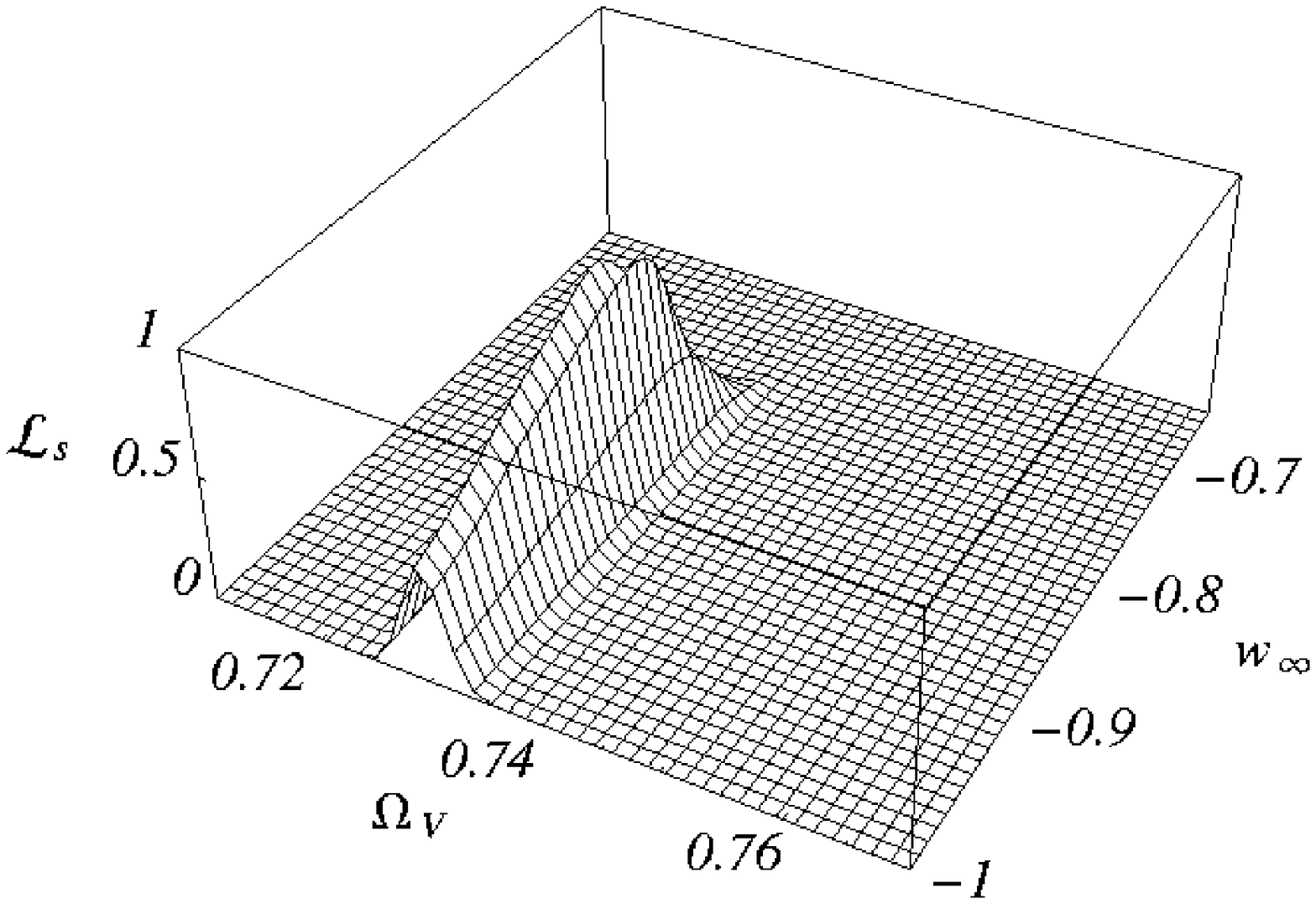}
    \includegraphics[clip=true]{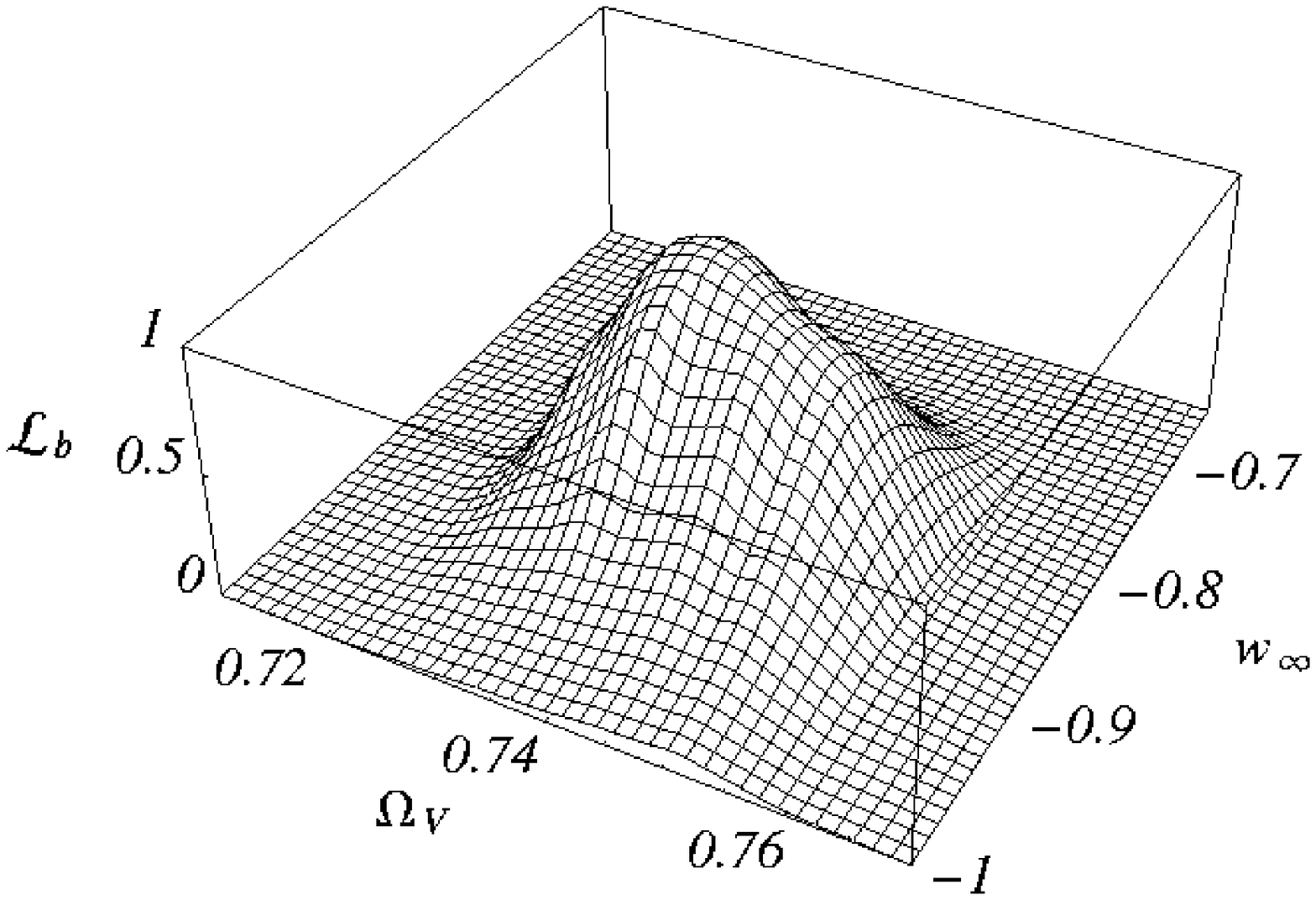}}
  \resizebox{\hsize}{!}{\includegraphics[clip=true]{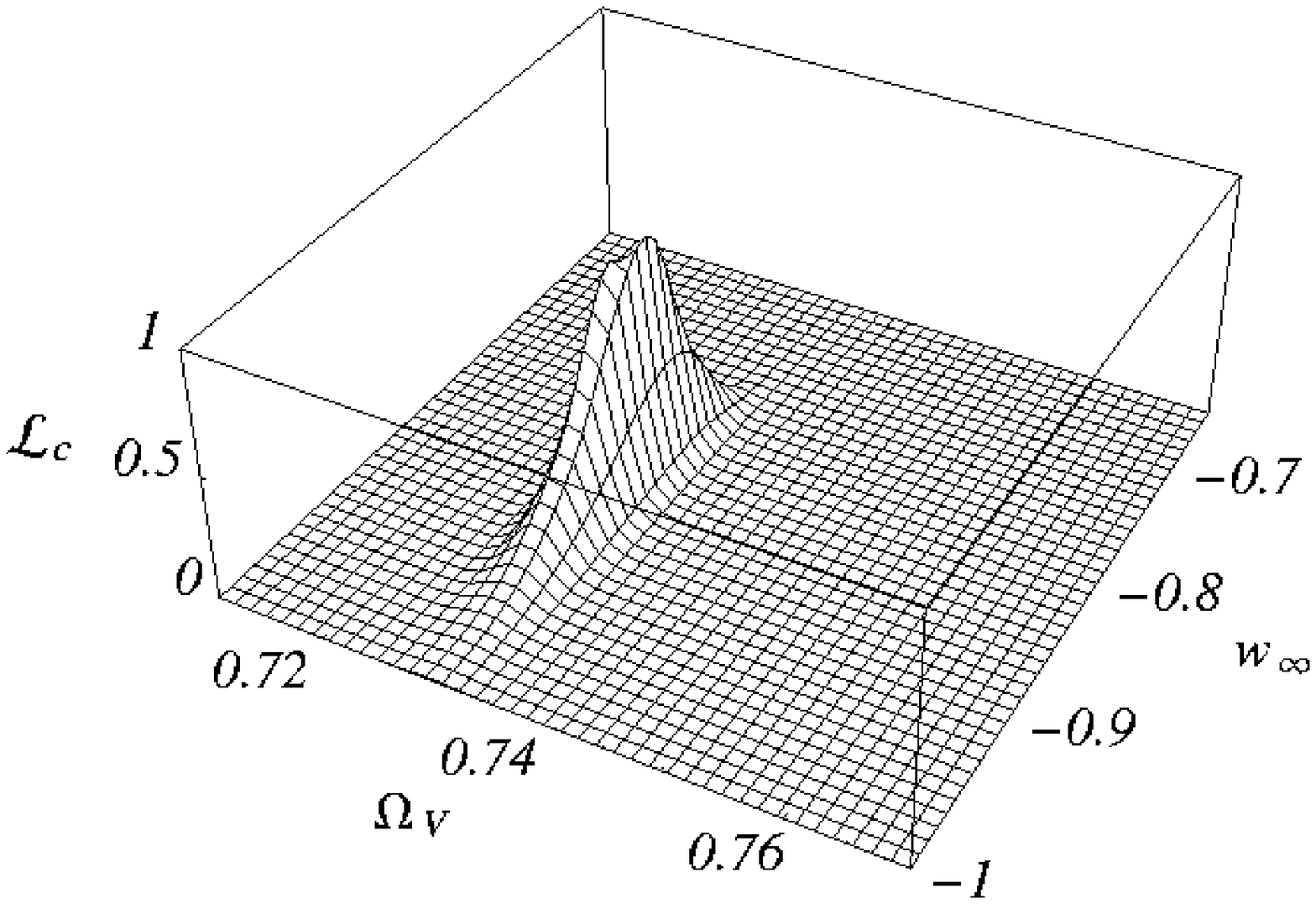}
    \includegraphics[clip=true]{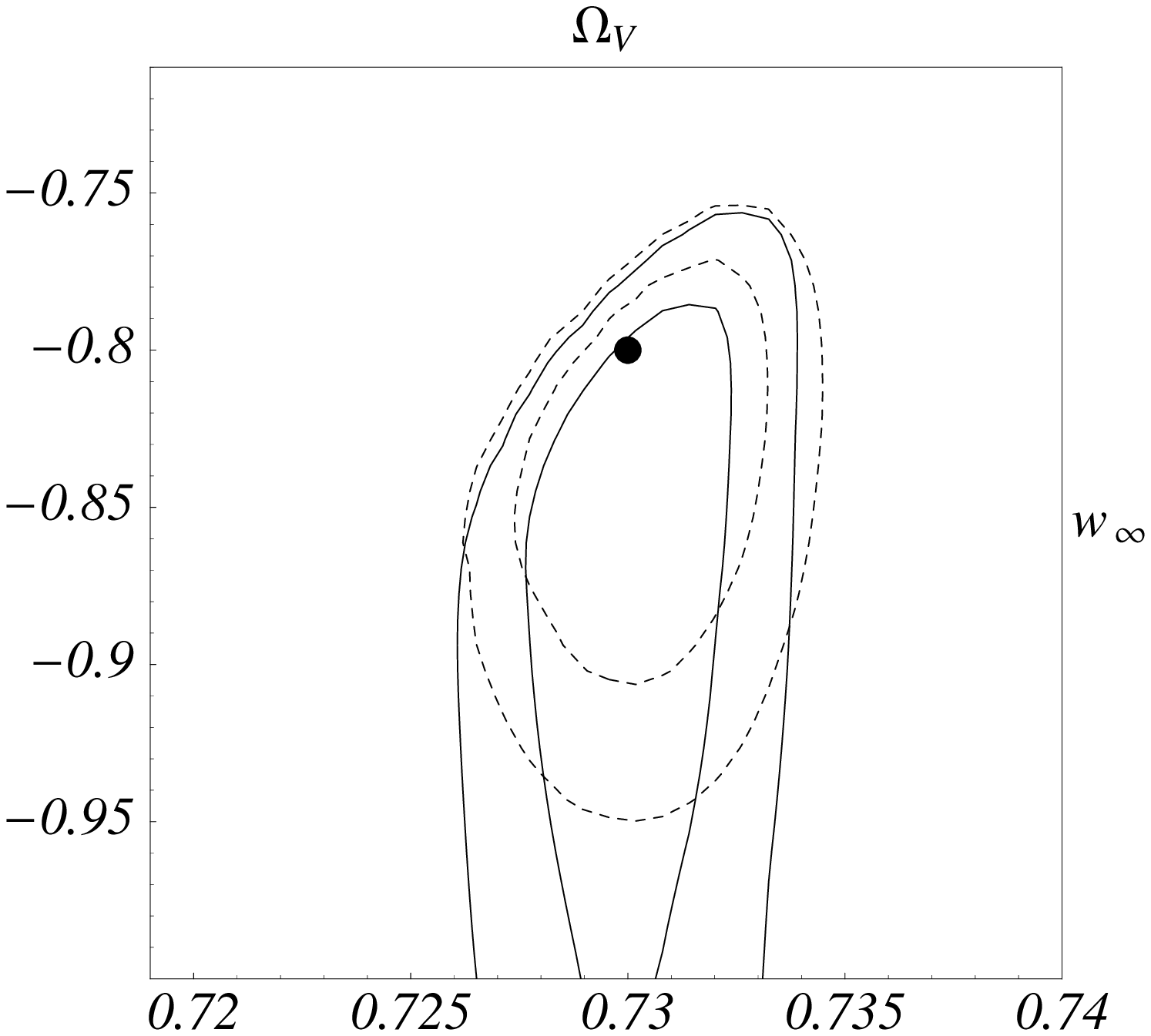}}
  \caption{As in Fig. \ref{f:w0wi_3d} but with marginalization over
    $w_0$.}
  \label{f:omegavwi_3d}
\end{figure}

\begin{figure}
  \centering
  \resizebox{\hsize}{!}{\includegraphics[clip=true]{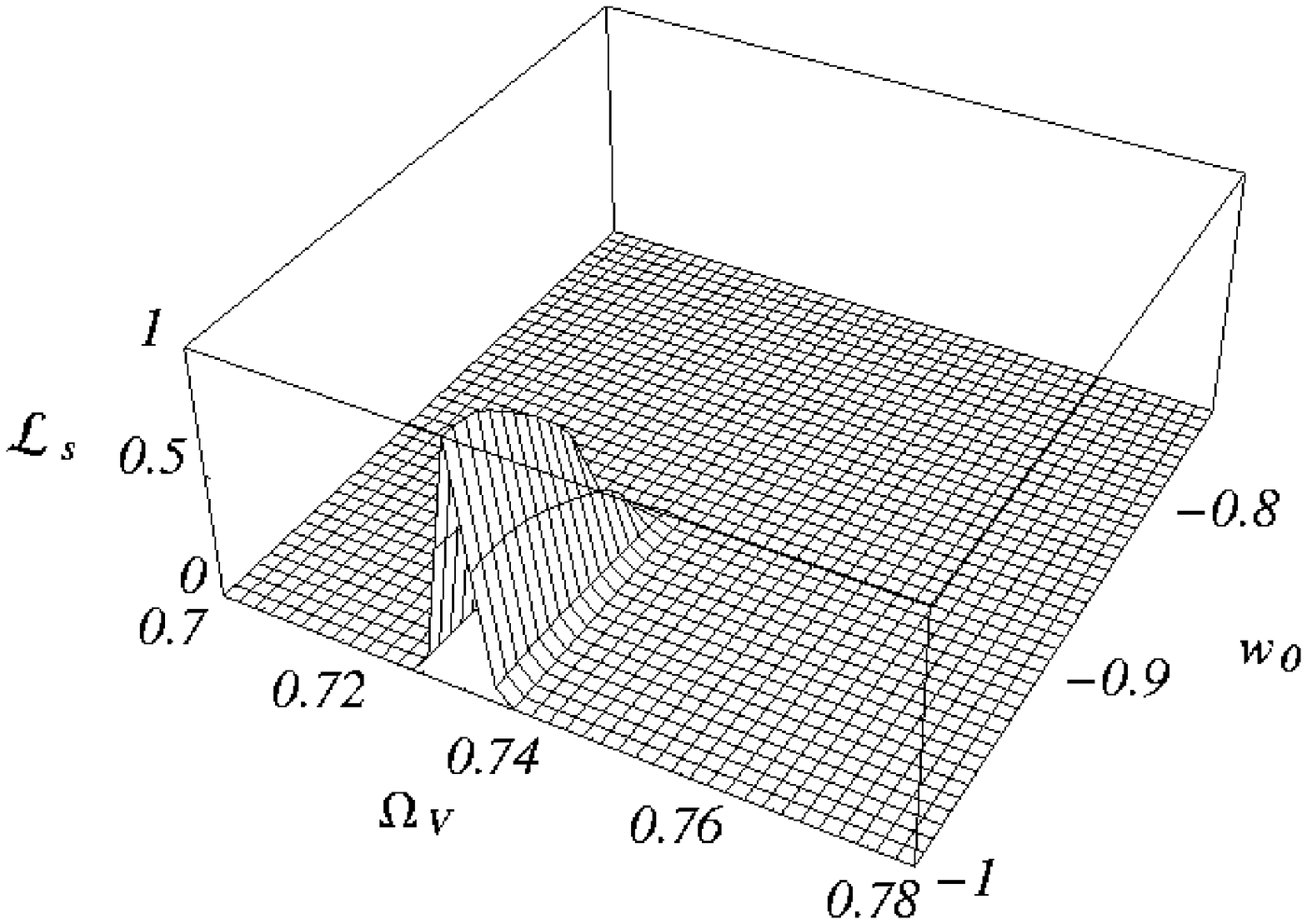}
    \includegraphics[clip=true]{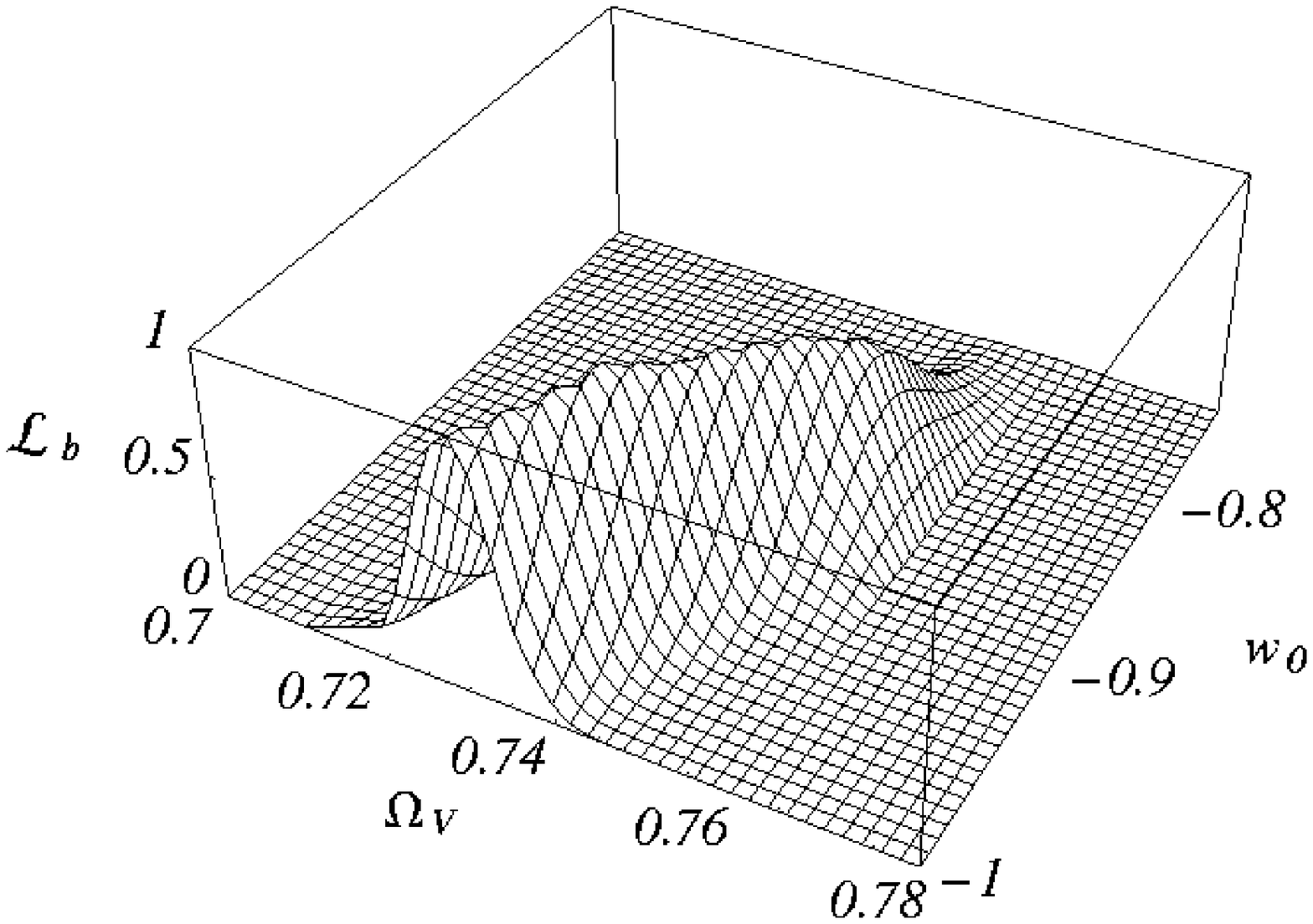}}
  \resizebox{\hsize}{!}{\includegraphics[clip=true]{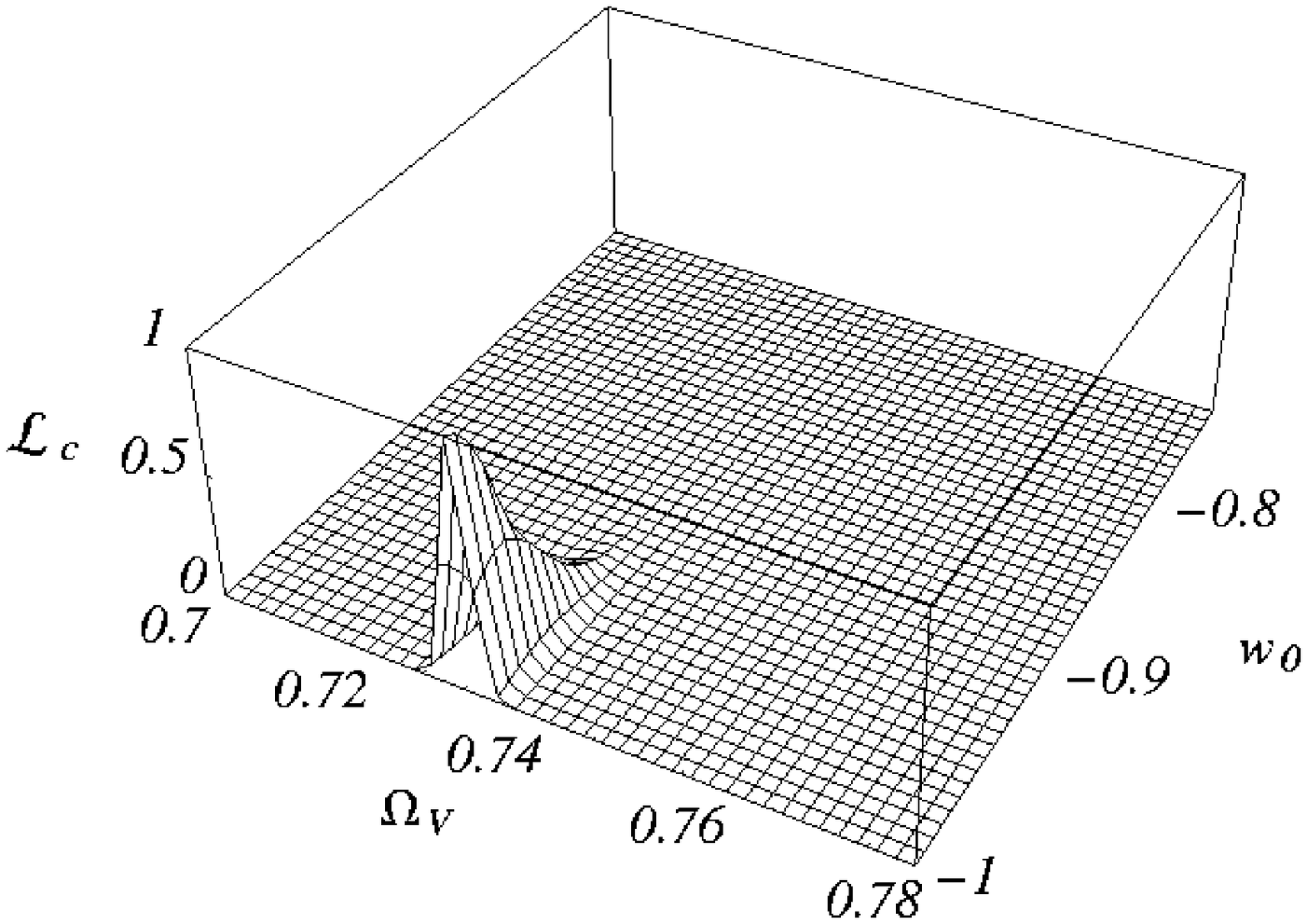} 
    \includegraphics[clip=true]{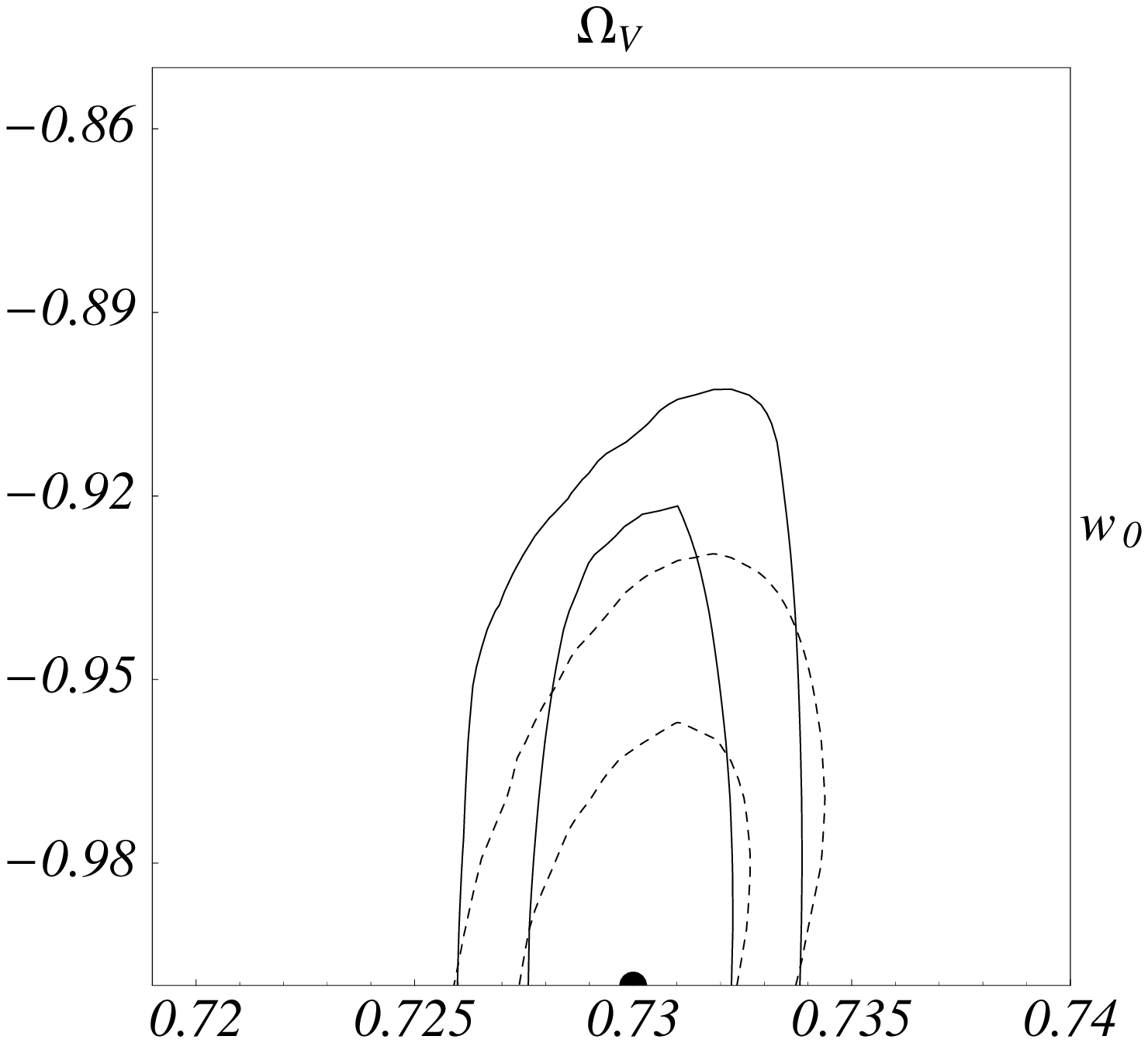}}
  \caption{As in Fig. \ref{f:w0wi_3d} but with marginalization over
    $w_\infty$.}
  \label{f:omegavw0_3d}
\end{figure}

In Fig. \ref{f:w0wi_3d}, \ref{f:omegavwi_3d} and \ref{f:omegavw0_3d}
we plot the three parameters likelihoods marginalized respectively
over $\Omega_V$, $w_0$ and $w_\infty$; in each figure, the top-left
panel is the power spectrum likelihood, the top-right panel is the
bispectrum likelihood, the bottom-left panel is the combined
likelihood and in the bottom-right panel we put the contour plots at
$2\sigma$ and $1\sigma$ confidence levels for the power spectrum
likelihood and the combined one.

As expected, when marginalizing over $\Omega_V$, the likelihood
analysis gives almost the same result as when $\Omega_V$ is fixed (see
Sec. \ref{two}). Indeed, the leading feature is the projection
degeneracy in the CMB power spectrum with respect to combined
variation of $w_0$ and $w_\infty$; the latter gets worse when also
$\Omega_V$ varies, although the latter variation affects also the
matter abundance inducing relevant changes in the power spectrum; the
resulting picture is therefore almost equivalent to the case when
$\Omega_V$ is fixed.

However, the three parameters case is different with respect to the
previous one when the marginalization is performed over $w_0$ or
$w_\infty$. In the previous sub-section we saw that when $w_0$ or
$w_\infty$ is fixed, the power spectrum constraints dominate the joint
likelihood. As we discussed above, having one of the two dark energy
equation of state parameters fixed means breaking the projection
degeneracy for the power spectrum, when also $\Omega_V$ varies. In the
present case the latter argument disappears, and the power spectrum
constraints are still fully affected by that degeneracy, causing the
visible gain from having the bispectrum into the analysis.

The top panels of Fig. \ref{f:omegavwi_3d}, obtained marginalizing
over $w_0$, reveal two different weaknesses of spectrum and
bispectrum, which make their combination fruitful. First, the weak
sensitivity of the power spectrum to $w_\infty$: the likelihood is
almost flat for a long line in the $w_\infty$ direction, until the
latter parameter induces a change strong enough to let the likelihood
vanishing. On the other hand, the sensitivity to $\Omega_V$ is strong
as its variation affects the matter abundance, inducing important
changes in the power spectrum shape. The opposite happens for the
bispectrum. Now the preferred parameter is $w_\infty$, as it
determines the dark energy behavior at the time when the lensing power
is injected. This orthogonality in the degeneracy direction of
spectrum and bispectrum determines the great advantage of their
combination, clearly visible in the bottom panels of
Fig. \ref{f:omegavwi_3d}.

The minimum gain is got in the case of Fig. \ref{f:omegavw0_3d}, where
the marginalization is made on $w_\infty$. Indeed, the latter
operation washes out the parameter of greatest importance for the
bispectrum; in the top-right panel, the bispectrum degeneracy
direction roughly corresponds to the constancy of the dark energy
abundance at the time in which the lensing injects its power. On the
other hand, as we stressed above the gain is still relevant and
visible in the bottom panels of Fig. \ref{f:omegavw0_3d}, as the three
parameters analysis makes the power spectrum projection degeneracy
fully effective.

\begin{figure}
  \centering
  \resizebox{\hsize}{!}{\includegraphics[clip=true]{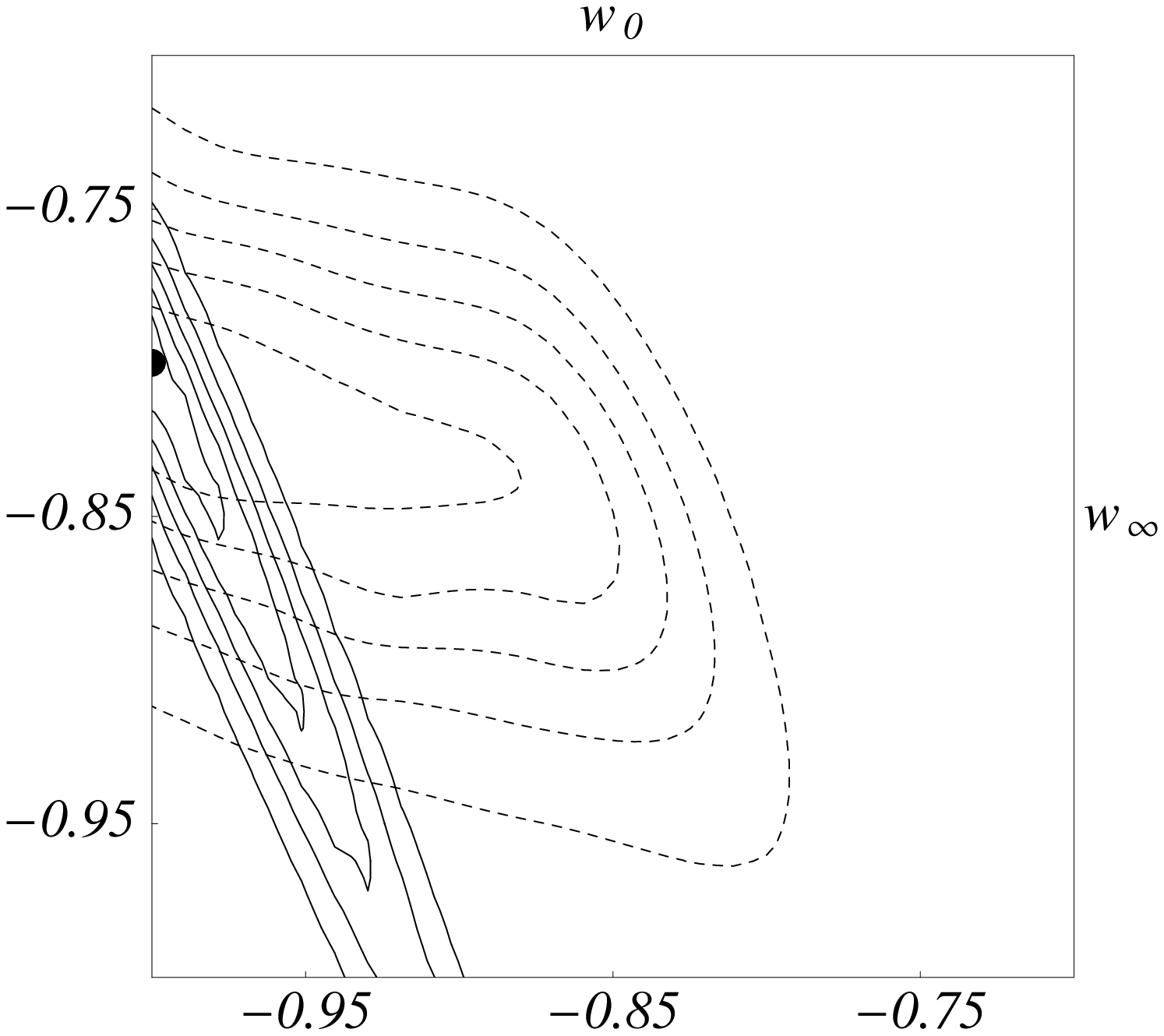}
    \includegraphics[clip=true]{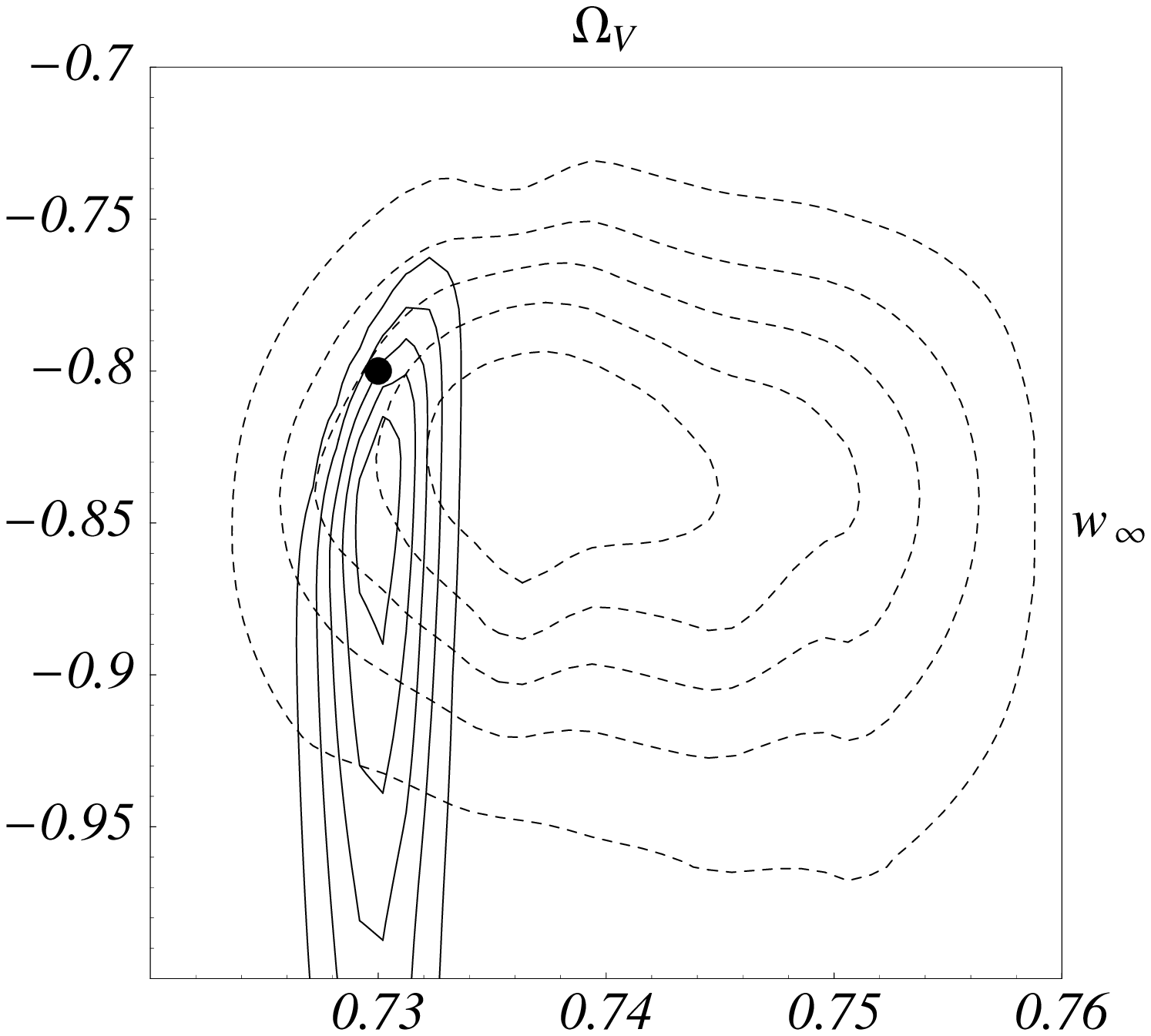}
    \includegraphics[clip=true]{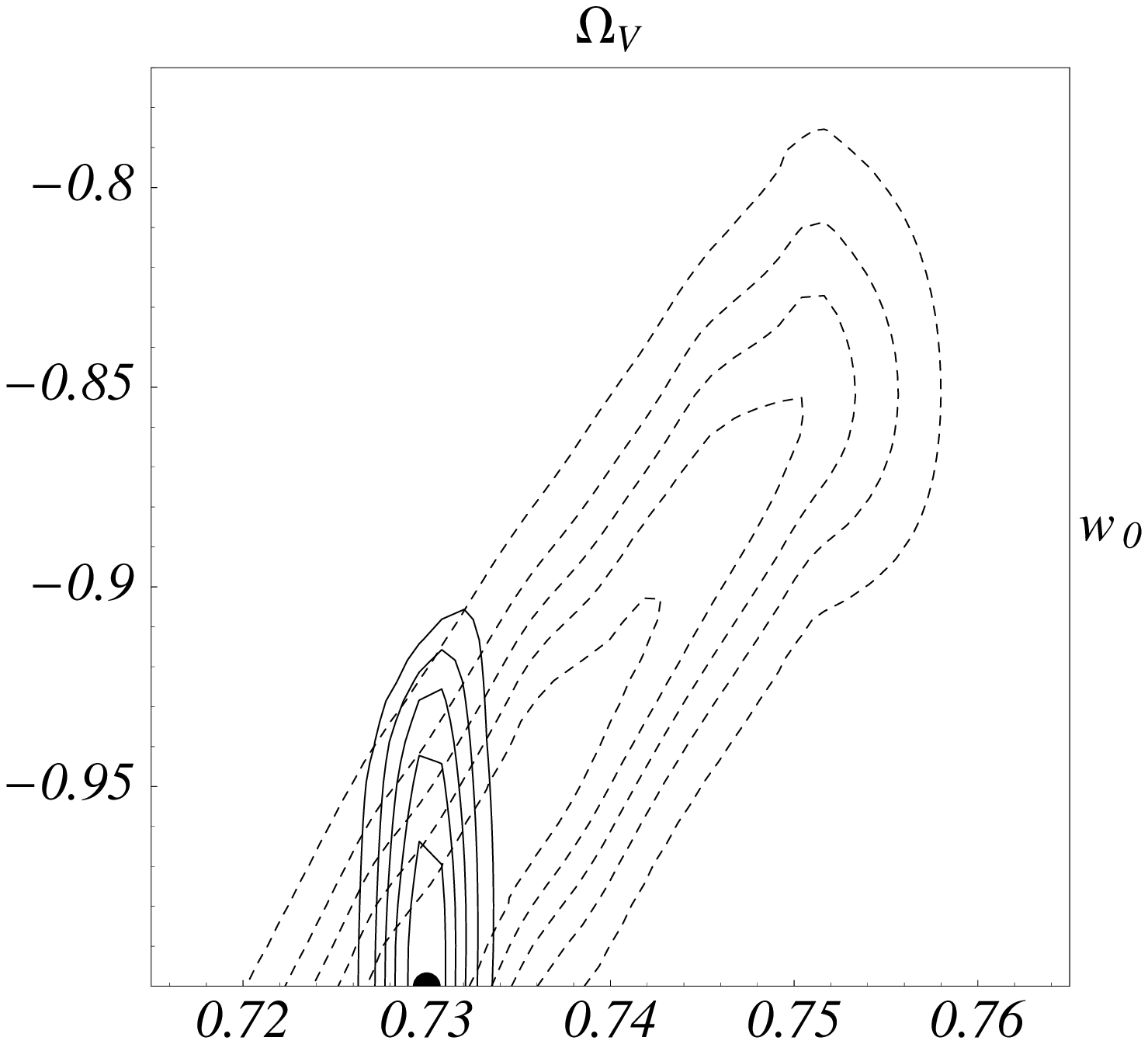}}
  \caption{Contour levels for the three parameters analysis, where one
    has been marginalized; the solid and dashed lines represent the
    power spectrum and bispectrum likelihoods, respectively; they are
    normalized to their maximum values and the five contours for each
    observable are equally spaced between 0 and 1 showing
    qualitatively the different basins of degeneracy. From left to
    right: marginalization over $\Omega_V$, marginalization over
    $w_\infty$, marginalization over $w_0$.}
  \label{f:degeneracy}
\end{figure}

In Fig. \ref{f:degeneracy} we give a qualitative picture of the
different likelihood shapes for spectrum and bispectrum, in the three
parameters analysis, where one has been marginalized; the solid and
dashed lines represent the power spectrum and bispectrum likelihoods,
respectively; they are normalized to their maximum values and the five
contours for each observable are equally spaced between 0 and 1 to
highlight their shape.

Finally, to quantify how the bispectrum analysis improves the 
estimation of the dark energy parameters, we report in Tab. I the 
constraints (both at $1\sigma$ and $2\sigma$) on our parameters, 
obtained when the power spectrum, bispectrum and the combined analysis 
are applied to our fiducial model. We stress that the other cosmological 
parameters are kept fixed. This table refers to the three parameter
likelihoods, marginalized over the remaining two of them. As
discussed, the confidence levels are always narrower when the
bispectrum analysis is added to the power spectrum one. Roughly, the
precision on the measures of $w_0$ and $w_\infty$ is percent and ten
percent, respectively.
\begin{table}
  \caption{Confidence level at $1\sigma$ and $2\sigma$ of
    marginalized three parameters likelihood.}
  \begin{center}
    \begin{tabular}{|c|c|c|c|}\hline
      $1\sigma$ c.l. & $\Omega_V$ & $w_0$ & $w_\infty$\\\hline
      Power spectrum  & $0.728 \le \Omega_V \le 0.732$ & $w_0 \le
      -0.95$ & $-0.94 \le w_\infty \le -0.81$\\
      Bispectrum & $0.731 \le \Omega_V \le 0.750$ & $w_0 \le -0.89$ &
      $-0.89 \le w_\infty \le -0.78$\\
      Combined & $0.729 \le \Omega_V \le 0.732$ & $w_0 \le -0.97$ &
      $-0.87 \le w_\infty \le -0.80$\\\hline
      $2\sigma$ c.l. & $\Omega_V$ & $w_0$ & $w_\infty$\\\hline
      Power spectrum & $0.727 \le \Omega_V \le 0.734$ & $w_0 \le
      -0.92$ & $w_\infty \le -0.79$ \\
      Bispectrum & $0.724 \le \Omega_V \le 0.759$ & $w_0 \le -0.79$ &
      $-0.96 \le w_\infty \le -0.74$\\
      Combined & $0.727 \le \Omega_V \le 0.734$ & $w_0 \le -0.94$ &
      $-0.93 \le w_\infty \le -0.77$\\\hline
    \end{tabular}
  \end{center}
  \label{t:confidence}
\end{table}

\section{CONCLUSIONS}
\label{conclusions}
We studied the non-Gaussian signal induced on the Cosmic Microwave
Background (CMB) total intensity anisotropies by weak gravitational
lensing, and its dependence on the redshift behavior of the
cosmological expansion rate. Our approach is based on the harmonic
expression of the three point CMB anisotropy statistics, which is
conveniently described in terms of the CMB bispectrum.

The lensing bispectrum signal depends on a triplet of angular
multipoles, connecting the primordial signal with the lensing kernel;
the latter is expressed through a redshift integral probing the
redshift derivative of the primordial power spectrum of density
perturbations on different scales projecting the same angle along the
line of sight. Such connection, appearing as a product of the
primordial signal and the lensing kernel on different multipoles,
determines a re-projection of the acoustic peaks on scales markedly
different from their well known location in the CMB power
spectrum. The bispectrum coefficients, seen as a function of the three
multipoles in arbitrary configuration, depict a complex structure of
peaks and valleys in a tri-dimensional space, bounded by geometrical
constraints encoded in the Wigner 3J symbols. A recurring feature is
the presence of a sign inversion in the bispectrum coefficients, on
angular scales of a few hundreds, cutting almost in the middle the
whole signal distribution; the latter feature represent the transition
between linear and non-linear power domination in the lensing kernel,
and appears visually as a ``canyon" in the distribution of the
absolute value of the bispectrum coefficients \cite{verde,giovi}.

Although rich, this phenomenology is unfortunately not observable in
detail, as the effect represents a second order cosmological
perturbation, and the single coefficient is largely dominated by
cosmic variance. The only way to exploit practically the bispectrum
coefficients is to compress the information by summing over the
different multipole triplets. By doing that, we verified that most of
the signal is contained in the triplets with different values of the
multipoles (``scalene" configurations), merely because they represent
the large majority of the whole number of coefficients. We evaluated
the signal to noise ratio by summing over all the triplets as a
function of the maximum multipole in the sum. In agreement with
previous analyses \cite{hu}, we show how a cosmic variance limited
experiment should be able to detect the bispectrum signal by summing
at least up to a maximum multipole of a few hundreds.

As it is known from previous works, the lensing signal is injected at
redshift higher than the present, and is therefore promising as a
probe of the expansion rate at that epoch. We parameterized the
redshift behavior of the expansion rate by means of the dark energy
abundance relative to the critical cosmological density today,
$\Omega_V$, its present and high redshift equation of state, $w_0$ and
$w_\infty$, respectively \cite{linder}. For CMB studies, the
expectation \cite{giovi} is that the bispectrum data may increase the
overall sensitivity of the CMB on the dark energy high redshift
dynamics, represented by $w_\infty$, in comparison with the usual
analysis made on the basis of the CMB total intensity power spectrum
only; the latter is sensitive to the redshift average of the dark
energy equation of state, through a projection effect: different
combinations of $w_0$ and $w_\infty$ leading to the same redshift
average cannot be distinguished.

To assess to what level such degeneracy is broken by taking into
account the CMB bispectrum, we set up a maximum likelihood analysis
simulating a Planck-like experiment, varying $\Omega_V$, $w_0$ and
$w_\infty$ and keeping all the other cosmological parameters fixed to
our fiducial model. Despite of the lower signal to noise ratio, the
bispectrum likelihood contours present a substantial misalignment with
respect to those of the power spectrum, being more sensitive to
changes in $w_\infty$; this is consistent with the expectation quoted
above, i.e. that the lensing should probe directly the expansion rate
at the epoch when the process is effective, independently on the
present. The bispectrum actually breaks the degeneracy of the CMB on
the redshift behavior of the dark energy, allowing a detection of both
$w_0$ and $w_\infty$, at least in our three parameters likelihood
analysis, where those and the dark energy abundance are the only
varying parameters; the level of accuracy is at percent and ten
percent, respectively. These results may increase the interest and
efforts toward the detection of the weak lensing signal by the
forthcoming CMB probes, as that may be relevant to gain insight into
the dark energy dynamics at the onset of cosmic acceleration, when
most models similar to a Cosmological Constant at present predict very
different behaviors.

A number of caveats or potential obstacles should be pointed out here
against this expectation.

Our analysis is based on a few parameters only, directly related to
the dark energy. Of course a multi-parameter study would possibly
reveal dangerous degeneracies in the bispectrum dependence on the
underlying cosmology. On the other hand, the CMB spectrum by itself,
as well as the wealth of independent cosmological observations are
fixing with high accuracy the main cosmological parameters. Such
accuracy will even increase with the observations by the forthcoming
probes. In this framework, it is reasonable to conceive focused
analyses based on specialized observables to probe critical aspects of
the whole cosmological picture. A study of this kind focused on the
dark energy dynamics would fit in this category: that would involve
the non-Gaussian CMB signal injected by the weak lensing and its
relation with the redshift behavior of the cosmological expansion
rate, but would take place within the confidence region allowed to the
the remaining cosmological parameters by the collection of
cosmological observations.

Another important aspect concerns directly the lensing kernel. As we
stressed above, a relevant part of the bispectrum signal comes from
the non-linear tail in the cosmological perturbation power
spectrum. The reliability of any constraint coming from this
observable depend crucially on the correct modeling of that part of
the spectrum. Although several recipes have been proposed and
currently used in many applications, this issue would deserve a
particular attention as a crucial piece of the constraining power of
the non-Gaussian CMB signal from lensing.

Also, the early universe itself may be a source of non-Gaussianity in
the primary CMB signal, and the weak lensing distortion should be
compared with the corresponding one from non-Gaussian models of
primordial perturbations.

Moreover, even if the cosmological uncertainties mentioned above are
under control, the foreground astrophysical signals may affect the
final result and need to be controlled. Due to the weakness of the
signal, both diffuse Galactic foregrounds and extra-Galactic point
sources or Sunyaev-Zel'Dovich (SZ) effects might be dangerous. A study
of the foreground non-Gaussian signal in comparison to the CMB one
from weak lensing will be certainly possible as the CMB observations
are providing an excellent improvement on our knowledge of the
foreground emission. Some very preliminary considerations may be
attempted concerning the first evidence of non-zero bispectrum signal
from radio sources \cite{wmap_ps}, as well as the predicted SZ signal
and the possible contribution from the early universe, which have been
reviewed recently \cite{bartolo}. Specifically, we evaluated the
amplitude of the lensing signal with respect to those in figure 14 in
the latter work; this indicates that it dominates over the radio
sources and SZ contributions, which are significant at high angular
scales, where we do not add more information by adding terms in the
sum in Eq. (6). On the other hand, the signal seems comparable with
the level of non-Gaussianity allowed by the present experiments in the
early universe \cite{wmap_ps}. This is a very naive indications that
the extra-Galactic foregrounds may not be dramatic as a contaminant of
the lensing signal; the same indications, extended to all foregrounds,
has been reported independently \cite{hirata}; that is supported by
the fact that some control of foregrounds may be achieved exploiting
their different scaling in frequency with respect to the CMB in
multi-band observations \cite{wmap_ps}. On the other hand, a major
contaminant, if present, may be represented by the primordial
non-Gaussianity, and one should exploit its different shape with
respect  to the lensing signal in order to detect both independently.

Finally, almost all the known instrumental systematics, including sky
cut, are a source on non-Gaussianity in the observed CMB
pattern. Although the future probes promise an excellent control of
systematics, the instrumental performance should be checked against
the cosmological signal also for what concerns higher order statistics
in the CMB pattern.

Despite of all these issues, and in the light of the results presented
above, it is certainly worth to address these caveats to assess their
relevance for CMB bispectrum measurements. Indeed, cosmic acceleration
constitutes a pillar of the whole cosmological picture, although not
understood. Any observable probing the transition to acceleration may
be crucial to spread light on it in the forthcoming years.

\section*{ACKNOWLEDGMENTS}
The authors acknowledge Scott Dodelson, Eric Linder, and Uros Seljak
for useful discussions. C.B. and F.G. are grateful to Davide Maino and
Roberto Trotta for their suggestions. F.G. would like to thanks
Michele Casula and Federico Gasparo for their computational support.
C.B. and F.P. were supported in part by by NASA LTSA grant
NNG04GC90G.

\bibliographystyle{aa}

\begin{thebibliography}{99}
\bibitem{lambda_review} P.J.E. Peebles and B. Ratra,
  Rev. Mod. Phys. 75, 599 (2003); T. Padmanabhan, Phys. Rep. 380, 235
  (2003).
\bibitem{de_models} L. Amendola, Phys. Rev. D69, 103524 (2004);
  S. Matarrese, C. Baccigalupi and F. Perrotta, Phys. Rev. D 70,
  061301(R), (2004).
\bibitem{linder} E.V. Linder, Phys. Rev. Lett., 90, 091301 (2003).
\bibitem{polarski} M. Chevallier and D. Polarski, Int. J. Mod. Phys. D
  10, 213 (2001).
\bibitem{sn1a} A.G. Riess et al., Astrophys. J. 116, 1009 (1998);
  S. Perlmutter et al., Astrophys. J. 517, 565 (1999).
\bibitem{wmap} D.N. Spergel et al., Astrophys. J. Suppl. Series 148, 175
  (2003).
\bibitem{tegmark} M. Tegmark et al., Phys. Rev. D69, 103501 (2004).
\bibitem{bacci} C. Baccigalupi, A. Balbi, S. Matarrese, F. Perrotta
  and N. Vittorio, Phys. Rev. D65, 063520 (2002).
\bibitem{perlsnap} S. Perlmutter, Nucl. Phys. B Proc. Suppl., 124, 13 (2003)
\bibitem{refregier04} A. Refregier et al., Astron.J. 127, 3102 (2004).
\bibitem{wl_review} M. Bartelmann and P. Schneider, Phys. Rep. 340,
  291 (2001); A. Refregier, Ann. Rev. Astron. Astrophys. 41, 645
  (2003).
\bibitem{hu} W. Hu, Phys. Rev. D 62, 043007 (2000).
\bibitem{acquaviva} V. Acquaviva, C. Baccigalupi and F. Perrotta,
  Phys. Rev. D 70, 023515 (2004).
\bibitem{hirata} C.M. Hirata, N. Padmanabhan, U. Seljak, D. Schlegel
  and J. Brinkmann, Phys. Rev. D70, 103501 (2004).
\bibitem{spergel_goldberg} D.N. Spergel and D.M. Goldberg,
  Phys. Rev. D 59, 103001 (1999); D.M. Goldberg and D.N. Spergel,
  Phys. Rev. D 59, 103002 (1999).
\bibitem{komatsu} E. Komatsu and D. N. Spergel, Phys. Rev. D63, 063002
  (2001). 
\bibitem{verde} L. Verde and D.N. Spergel, Phys. Rev. D 65, 043007
  (2002).
\bibitem{giovi} F. Giovi, C. Baccigalupi and F. Perrotta, Phys. Rev. D
  68, 123002 (2003).
\bibitem{corasaniti03} P.S. Corasaniti and E.J. Copeland,
  Phys. Rev. D67, 063521, (2003).
\bibitem{corasaniti04} P.S. Corasaniti, M. Kunz, D. Parkinson,
  E.J. Copeland and B.A. Bassett, Phys. Rev. D70, 083006 (2004);
  H.K. Jassal, J.S. Bagla and T. Padmanabhan, preprint
  astro-ph/0404378, to appear in Mon. Not. R. Astron. Soc. Lett.,
  (2004); B. Gold, preprint astro-ph/0411376, submitted to
  Phys. Rev. D (2004).
\bibitem{ma} C.P. Ma, Astrophys. J. Lett. 508, L5, (1998); C.P. Ma,
  R.R. Caldwell, P. Bode and L. Wang, Astrophys. J. Lett 521, L1
  (1999).
\bibitem{afs} N. Afshordi, Y.S. Loh and M.A. Strauss, Phys. Rev. D69,
  083524 (2004).
\bibitem{balbi} A. Balbi, C. Baccigalupi, F. Perrotta, S. Matarrese
  and N. Vittorio, Astrophys. J. Lett. 588, L5 (2003).
\bibitem{gangui} Gangui A. and Martin J.,
  Mon. Not. R. Astron. Soc. 313, 323 (2000).
\bibitem{takada} M. Takada and B. Jain, Mon. Not. R. Astron. Soc. 348,
  897 (2004).
\bibitem{wmap_ps} Komatsu et al., Astrophys. J. Supp. Series, 148,
  119, (2003).
\bibitem{bartolo} Bartolo N., Komatsu E., Matarrese S., Riotto A.,
  Phys. Rep. 402, 103 (2004).
\end{thebibliography}

\end{document}